\documentclass[%
 reprint,
superscriptaddress,
 amsmath,amssymb,
 aps,
]{revtex4-2}

\usepackage{graphicx}
\usepackage{dcolumn}
\usepackage{bm}
\usepackage{enumitem}
\usepackage{braket}
\usepackage{url}
\usepackage{hyperref}
\hypersetup{colorlinks=true, citecolor=blue, urlcolor=blue, linkcolor=blue, unicode=true}

\usepackage{txfonts}

\begin{document}

\preprint{APS/123-QED}

\title{Digital-analog quantum genetic algorithm using Rydberg-atom arrays}

\author{Aleix Llenas}
\affiliation{Universidad Internacional Men\'endez Pelayo, 28040 Madrid, Spain}

\author{Lucas Lamata}
\affiliation{Departamento de F\'isica At\'omica, Molecular y Nuclear, Facultad de F\'isica, Universidad de Sevilla, Apartado 1065, E-41080 Sevilla, Spain}

\date{\today}

\begin{abstract}

Digital-analog quantum computing (DAQC) combines digital gates with analog operations, offering an alternative paradigm for universal quantum computation. This approach leverages the higher fidelities of analog operations and the flexibility of local single-qubit gates. In this paper, we propose a quantum genetic algorithm within the DAQC framework using a Rydberg-atom emulator. The algorithm employs single-qubit operations in the digital domain and a global driving interaction based on the Rydberg Hamiltonian in the analog domain. We evaluate the algorithm performance by estimating the ground-state energy of Hamiltonians, with a focus on molecules such as $\rm H_2$, $\rm LiH$, and $\rm BeH_2$. Our results show energy estimations with less than 1\% error and state overlaps nearing 1, with computation times ranging from a few minutes for $\rm H_2$ (2-qubit circuits) to one to two days for $\rm LiH$ and $\rm BeH_2$ (6-qubit circuits). The gate fidelities of global analog operations further underscore DAQC as a promising quantum computing strategy in the noisy intermediate-scale quantum era.

\end{abstract}

\maketitle


\section{\label{sec:level1} Introduction}

Quantum computing is poised to address significant challenges across various domains, particularly in many-body physics, where it can determine the ground state of strongly interacting quantum systems \cite{RevModPhys.86.153, mcclean2016theory}. This capability is especially relevant in quantum chemistry \cite{aspuru2005simulated}, where accurately finding the ground state energies of complex molecules is crucial for various applications including combustion \cite{gonthier2022measurements}, batteries \cite{kim2022fault, delgado2022simulating}, and catalysts \cite{goings2022reliably}, and where quantum simulations hold great promise.

Recent technological advances indicate that we are entering a new era of quantum computing, one in which classical computers struggle to simulate the behavior of programmable quantum computers \cite{arute2019quantum}. In this context, it is anticipated that the first practically useful algorithms will be heuristic.

Quantum computing is generally divided into two paradigms: digital quantum computing (DQC) \cite{deutsch1989quantum, yao1993quantum} and analog quantum computing (AQC) \cite{johnson_what_2014}. DQC performs computations using discrete quantum gates, analogous to classical digital circuits. It relies on qubits manipulated through well-defined operations such as single-qubit gates (like rotations) and two-qubit gates [like controlled-NOT (CNOT)]. While DQC emphasizes error correction and fault tolerance, implementing these features demands significant physical resources.

In contrast, AQC operates with continuous variables, utilizing the natural dynamics of quantum systems. Instead of discrete gates, AQC employs continuous-time evolution governed by the system’s Hamiltonian. This approach is inherently robust against certain types of noise and decoherence \cite{greiner2002quantum}, making it an attractive alternative.

A third paradigm, the digital-analog quantum computing (DAQC) paradigm, has recently been proposed \cite{lamata, parra2020digital}. DAQC combines the strengths of both digital and analog quantum computing. DAQC integrates the flexibility of digital gates with the robustness of analog entangling operations. Typically, DAQC involves single-qubit operations within the digital domain, and analog multi-qubit entangling operations leveraging the natural interaction Hamiltonian of the system in use. Common quantum algorithms like the Quantum Fourier Transform \cite{martin2020digital} and the Quantum Approximate Optimization Algorithm \cite{headley2022approximating} have been successfully adapted to this paradigm.

DAQC can be implemented within the limitations of current noisy intermediate-scale quantum (NISQ) devices, with examples found in Rydberg atoms, trapped ions, and superconducting circuits \cite{daley2022practical, joshi2024observing, gong2023quantum, andersen2024thermalization, lu2024digital}.

Rydberg quantum simulators represent a significant advancement in quantum simulation platforms, attributed to their remarkable scalability and programmability \cite{henriet2020quantum, bluvstein2024logical}. These simulators operate by trapping individual atoms within optical tweezer arrays, which can interact when elevated to Rydberg states. This platform inherently executes spin Hamiltonians and has achieved analog quantum simulation with hundreds of atoms \cite{scholl2021quantum}. Key features of Rydberg-atom simulators include their ability to position atoms arbitrarily in two and three dimensions, offering substantial flexibility in connectivity, and their capacity to prepare various initial product states as heuristic trials prior to unitary evolution, either through digital gates or analog Hamiltonian evolution.

Despite the focus of most quantum computing research on the digital paradigm, with notable open-source software like Qiskit \cite{qiskit2024}, Cirq \cite{cirq_developers_2024_11398048}, and Pennylane \cite{bergholm2022pennylane}, fewer developments have been made in the analog paradigm. Previous papers that explored the DAQC paradigm followed a more hands-on approach using the package Pulser \cite{Silverio2022pulseropensource}, an open-source package for the design of pulse sequences in programmable neutral-atom arrays. However, Pasqal \cite{noauthor_pasqal_nodate} has recently released an open-source package specifically designed for the DAQC paradigm: Qadence \cite{qadence2024pasqal}. Qadence abstracts the complexity of low-level pulse programming, providing a high-level interface for building digital-analog quantum programs. 

To this end, Qadence works with the backend Pulser and thus one can convert circuits to the equivalent squared laser pulses to control Rydberg-atom arrays. This creates a direct connection between high-level Python-made digital-analog programs and low-level laser pulse sequences for execution on real devices. In this paper, we have leveraged Qadence to develop a digital-analog quantum genetic algorithm. 

Even though we have no current access to a scalable Rydberg quantum platform, it is expected that in 2-3 years Rydberg devices with several hundred atoms will be accessible via, e.g., online services such as the ones provided by the companies Pasqal \cite{noauthor_pasqal_nodate} and QuEra \cite{quera}, enabling the practical testing of this proposal.

Genetic Algorithms (GA) are heuristic optimization algorithms that were first developed in the 1960s by Bremermann and Holland \cite{bremermann1958evolution, 10.1145/321127.321128}. Today, GAs are widely used in various optimization fields, including price optimization, scheduling applications, and vehicle routing problems \cite{Reeves2010}. GAs offer several advantages over other optimization algorithms:

\begin{enumerate}[label=(\roman*), itemsep=0pt, parsep=0pt]
    \item Parallelism: GAs explore a population of solutions simultaneously, rather than a single solution, potentially enhancing the speed of the optimization process.
    \item Adaptability: By investigating multiple pathways concurrently, GAs increase adaptability and are more likely to find global optima, reducing the risk of getting trapped in local optima.
\end{enumerate}

However, GAs also present challenges. Implementing GAs requires careful design of the objective function and proper selection of representation and operators, making the process somewhat artisanal. Additionally, GAs can be more computationally expensive and time consuming than other methods.

The typical structure of a genetic algorithm involves the following steps:

\begin{enumerate}[label=(\roman*), itemsep=0pt, parsep=0pt]
    \item Initialization: Start with a randomly generated population of solutions.
    \item Selection: Evaluate and rank solutions based on a fitness function, discarding the worst half.
    \item Crossover: Combine pairs of high-ranking solutions to create new solutions, filling the discarded half of the population.
    \item Mutation: Apply random mutations to some solutions.
    \item Iteration: Repeat the selection, crossover, and mutation steps until a stopping criterion is met.
\end{enumerate}

The specifics of the crossover and mutation processes can vary depending on the problem and the nature of the solutions. Despite these variations, the fundamental structure of GAs—initial population, selection, crossover, and mutation—remains consistent.

Quantum versions of genetic algorithms have primarily been explored within the digital quantum computing paradigm, with few proposals in the analog paradigm. This paper proposes a quantum genetic algorithm within the DAQC framework.

Previous studies in the digital paradigm have proposed crossovers based on replicating selected individuals using an approximate quantum cloning machine and then combining their features with qubit swap operations. Mutations were implemented using random single-qubit rotations \cite{10186457}. 

There is also an analog approach, known as the Quantum-Assisted Genetic Algorithm, using quantum fluctuations for mutations and classical mechanisms for crossovers that was tested on a D-Wave \cite{noauthor_d-wave_nodate} 2000Q quantum annealing processor \cite{king2019quantumassisted}.

Our paper builds on related efforts by Antoine Michel et al., who developed a variational quantum eigensolver algorithm to find the ground state energy of Hamiltonians derived from chemistry \cite{michel2023blueprint}. They used a classical optimization algorithm in conjunction with the Rydberg-atom topological register, converting their algorithm into laser pulses with the Pulser backend. Similarly, Jonathan Z. Lu et al. proposed a variational quantum algorithm using a classical optimizer to perform digital-analog learning tasks on Rydberg-atom processors, providing a deep numerical study of various models \cite{lu2024digital}.

In this paper, we introduce a hybrid quantum genetic algorithm developed under the DAQC paradigm, simulating it on a Rydberg-atom array emulator. We evaluate the algorithm's performance using Hamiltonians coming from chemistry.

\section{Methods}
We describe below the operations performed, the protocol designed and the method to derive those Hamiltonians.

\subsection{Digital-Analog operations on Rydberg atom arrays}

Qadence allows us to construct quantum circuits that can be directly executed in Rydberg-atom processors thanks to the Pulser backend. However, we have to restrict the number of operations to single-qubit gates and global analog rotations \cite{qadence2024pasqal}.

Qadence translates digital-analog programs to a pre-defined system Hamiltonian representing Rydberg-atom arrays. In these devices, atoms can be arranged in arbitrarily shaped register layouts, and computations are realized by irradiating the array with appropriately tuned laser pulses. During the computation, qubits evolve under an effective Hamiltonian as defined by Eq.~(\ref{eq:one}):

\begin{eqnarray}
H = \sum_{i=0}^{n-1} \left( H_i^d + \sum_{j<i} H_{ij}^{int} \right)    \quad  ,
\label{eq:one}
\\
H_i^d = \frac{\Omega}{2} [cos(\phi) X_i - sin(\phi) Y_i ] - \delta N_i   \quad ,
\label{eq:two}
\\
H_{ij}^{int} = \frac{C_6}{r_{ij}^{6}} N_i N_j    \quad .
\label{eq:three}
\end{eqnarray}

Eq. (\ref{eq:two}) and (\ref{eq:three}) show the driving and interaction terms, where the Rabi frequency $\Omega$ , detuning $\delta$, and phase $\phi$ are parameters describing global laser pulses. $N_i = (I - \sigma_i^z)/2$  is the number operator to describe state occupancy. The interaction strength scales with $C_6$, a coefficient dependent on the quantum number in which the atomic array has been prepared, and decays with the distance between the atoms $r_{ij}^{6}$. For more information on quantum computing with neutral atoms we refer to \cite{henriet2020quantum} and \cite{bluvstein2024logical}.

For a given register of atoms prepared in some spatial coordinates, the Hamiltonians described will generate the dynamics of some unitary operation as Eq.~(\ref{eq:four}), where we specify the final parameter $t$, the duration of the operation:

\begin{eqnarray}
U(t, \Omega,\delta, \phi) = e^{-iHt} \quad .
\label{eq:four}
\end{eqnarray}

In this paper we have used the Qadence version v1.4.1 \cite{qadence2024pasqal}. The most general operation can be performed with the $AnalogRot$ function. This function allows the user to control all the relevant parameters:
\begin{verbatim}
  da_rot= AnalogRot(omega=’om’, phase=’ph’,
                    delta=’d’, duration=’t’)
\end{verbatim}

Qadence also allows one to control the distance between atoms, with a standard separation of 8 $\mu m$. This is a typical value used in combination with a standard experimental setup of neutral atoms such that the interaction term in the Hamiltonian can effectively be used for computations, and is the value used in this paper.

\subsection{Protocol for a Digital-Analog Quantum Genetic Algorithm}

The method developed in this paper follows the same structure as a genetic algorithm as described in the previous section.

\begin{figure}
\includegraphics[width=0.35\textwidth]{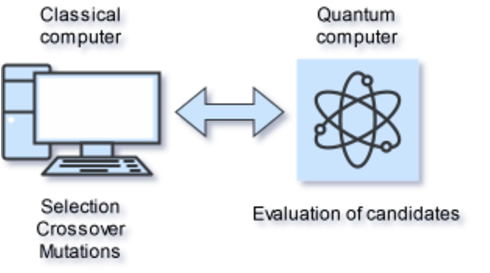}
\caption{\label{fig:schema} Schematic of the protocol: selection of candidates, crossover and mutations happens in the classical computer, while the evaluation of candidates happens in the quantum computer.}
\end{figure}

This is a hybrid algorithm: the selection of candidates, the crossover, and random mutations happen in a classical computer, and the candidates to be evaluated are sent to a quantum computer. There is a constant communication between both as described in Fig.~\ref{fig:schema}.

\subsubsection{Initial set of candidate solutions}

The protocol starts from a population of candidate solutions: each candidate solution consists of multiple qubits, depending on the problem to tackle. For the case of the $\rm H_2$ molecule, the problem Hamiltonian is expressed with only 2 qubits, and our candidates will be made of 2 qubits each, while if we want to find the ground state of the $\rm LiH$ or $\rm BeH_2$ molecules its problem Hamiltonian is expressed with 6 qubits, and we will need candidates with 6 qubits each. 

In our code we start with an initial population of 200 candidates. The circuit of each candidate is structurally the same for all of them but with different random parameters. 
With the set of available operations to be performed in Rydberg-atom arrays (see previous section), we can construct our ansatz in the following way: a combination of single qubit rotations (RX, RY and RZ) and two global interacting Hamiltonians.

From candidate to candidate the structure of the ansatz does not change, but the parameters do. Therefore, each of the 200 candidate solutions will have different values for these parameters: 

\begin{enumerate}[label=(\roman*), itemsep=0pt, parsep=0pt]
     \item RX: the rotation angle for each qubit.
     \item RY: the rotation angle for each qubit.
     \item RZ: the rotation angle for each qubit, but same on both sides of the analog block.
     \item First global analog block: does not change. We use a fixed set of parameters $(t,\Omega,\delta,\phi)$ for all candidates, not changing them at all.
     \item Second global analog block: fix set of parameters for $\Omega$, $\delta$ and $\phi$, only changing from candidate to candidate the evolution time $t$.
 \end{enumerate}

\begin{figure*}[btp!]
\includegraphics[width=\textwidth]{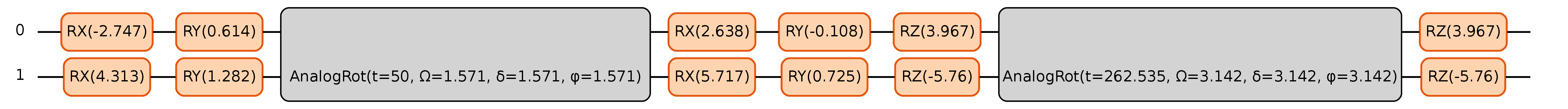}
\caption{\label{fig:h2_circuit_15} Two qubit ansatz used for the H2 molecule. The ansatz is composed of RX, RY and RZ single qubit rotations and two global interacting Hamiltonians.}
\end{figure*}

An example for a 2 qubit ansatz is shown in Fig. (\ref{fig:h2_circuit_15}).

The $\Omega,\delta,\phi$ parameters in both global analog blocks are not changed at all, and will be the same for all candidates during the whole protocol and equal to $\pi/2$ in the first block and equal to $\pi$ in the second block. In the first block, the time $t$ is also the same for all candidates (50 $ns$) and constant during the whole protocol. What does change is the evolution time $t$ for the second global analog block, allowing it to change for each candidate from iteration to iteration, and from candidate to candidate. 

The reason to justify this decision is that the first global analog block is introduced here to generate entanglement and diversity in the solution space. By sandwiching it between single qubit rotations that can take any value, we are making sure that regardless of the set of parameters $t, \Omega,\delta,\phi$ chosen, tweaking the RX and RY rotation angles in each qubit individually we can generate a good set of diverse candidates. In a purely digital setting, this would be equivalent to adding CNOT gates connecting all the qubits together. Notice that we are not aiming for universality here, but for sufficient diversity. 

In addition, in the second global analog block, that is also sandwiched between RZ rotation angles, we do allow the algorithm to play with the time $t$ because the whole combination not only gives more diversity and entanglement but also allows the protocol to converge towards the ground state of the problem Hamiltonian.

Once the 200 candidates with random configurations but the same structure are created, we perform an expectation value calculation with the problem Hamiltonian to all of them individually. The candidates and the expectation value calculation part $E = \braket{\psi|H_T|\psi}$ are executed in the quantum computer simulator. This is a state-vector simulator that does not consider noise in the gates. When we receive the results the candidates are sorted from lowest expectation values to highest, discarding the second half of the list (the worst solutions).

\subsubsection{Crossover}

Then, we perform the crossover with the first half of the list: by grouping the solutions from top to bottom in groups of two (parents), we create two child candidates with combined parameters taken from both parents. With this method we can now fill the second half of the list with new candidates that are originated with configurations from the first half of the list. 

\begin{figure*}[hbt!]
\includegraphics[width=0.85\textwidth]{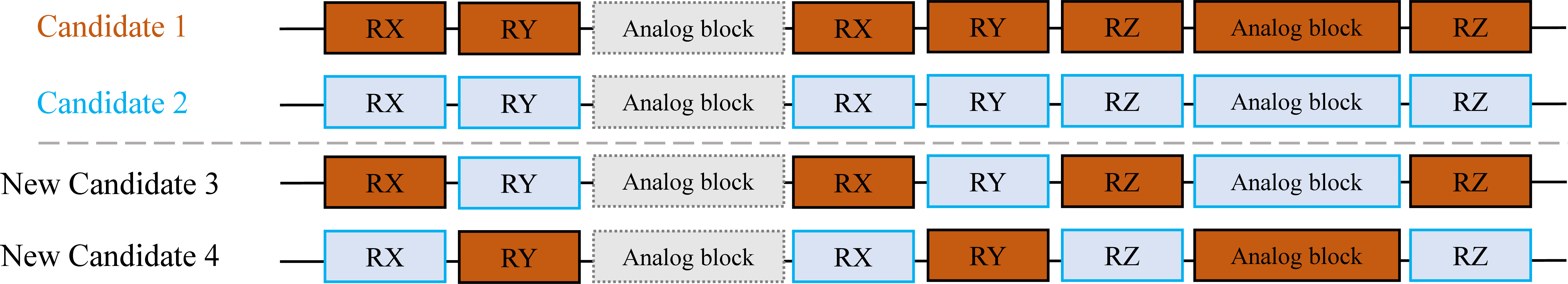}
\caption{\label{fig:crossover} Schematic of the crossover: two parent candidates (Candidate 1 and Candidate 2) are combined to generate two new solutions, New Candidate 3 and New Candidate 4. For the sake of simplification, the horizontal solid black line represents multiple qubits, and the different gates are applied to all of those multiple qubits.}
\vspace{-1ex}%
\end{figure*}

The crossover is performed by combining configurations from two parent candidates to create two new solutions. Since the first analog block has constant parameters, it is not modified and is the same for all candidates. In the second analog block only the evolution time $t$ is interchanged. In the single qubit rotations, the rotation angle is interchanged between candidates. A schematic of the process can be seen in the Fig. \ref{fig:crossover}. 

\subsubsection{Mutations}

The following step involves the mutations. To avoid falling into a local minimum and to provide more diversity into the solution space we perform some mutations to some qubits at random. These mutations are random rotations to the single-qubit gates and time modifications to the time evolution of the second analog block. To facilitate a convergence towards a final solution, for iteration after iteration the strength of the random change is reduced and the likelihood that a mutation can occur also gets lowered. Therefore, by the last steps of the protocol, when we are almost reaching the final solution, the number of mutations is less and less powerful (rotating angle smaller, change in evolution time smaller) than at the beginning of the algorithm. 

Then, we perform the sorting and selection of candidates by computing the expectation value with the problem Hamiltonian as described before and the protocol starts again.

The final stopping criteria can be selected in many different ways. In our testing, and since we can compute the exact diagonalization for those problem Hamiltonians, we set our stopping criteria when the energy difference between our solution and the exact ground state was less than 1\% as defined by Eq. (\ref{eq:six}):

\begin{eqnarray}
Error = \frac{E_{reference} - E_{found}}{E_{reference}} \quad .
\label{eq:six}
\end{eqnarray}

For real applications where the exact Hamiltonian is too complex and cannot be diagonalized classically, the stopping criteria could be set in different ways, for example, when some finite number of iterations have passed (time limit), or when the new energy found has only changed a very small percentage with respect to previous iterations (rate change limit), or a combination of the two. 

Since these problem Hamiltonians can be exactly diagonalized, we can perform an extra test to the protocol by computing the ground state overlap $|\braket{\psi|\Psi}|$, where $\Psi$ is the exact ground state eigenvector obtained from the diagonalization of the Hamiltonian and $\psi$ is the best solution found by the protocol, giving the operation 1 when both vectors are the same.

\subsection{Hamiltonians from Quantum Computational Chemistry}

In this section, we explain the process to derive the Hamiltonian of the above-mentioned molecules. We followed the same procedure as described in \cite{michel2023blueprint}. For more information we also refer to \cite{graves2023electronic}.

We begin by expressing the electronic Hamiltonian as a spin model \cite{PhysRevX.8.031022}. This approach originates from the Born-Oppenheimer approximation, which treats the nuclei of molecules as classical point charges (Eq. (\ref{eq:eight})):

\begin{eqnarray}
H = - \sum_i \frac{\nabla_i^2}{2} - \sum_{i,I} \frac{Z_i}{|r_i - R_I|} +\frac{1}{2} \sum_{i \neq j} \frac{1}{| r_i - r_j |} \quad ,
\label{eq:eight}
\end{eqnarray}

where $\nabla_i$ is the kinetic energy term for the $i-th$ electron, $Z_i$ is the charge of the $I-th$ nucleus, and $r,R$ denote the distance of the $i-th$ electron and the $I-th$ nucleus with respect to the center of mass, respectively. We aim to obtain the ground state energy of this Hamiltonian.

To represent electronic wavefunctions, a suitable basis set must be defined. We focus on the Slater-type orbital approximation for the basis set, specifically using three Gaussian functions. This minimal basis set includes the essential orbitals necessary to represent an atom valence shell. Additionally, the wavefunctions must exhibit antisymmetry under electron exchange. Achieving this requires second quantization, where anticommuting fermionic creation and annihilation operators are defined as $\{a_p^\dag\},\{a_p\}$ and rewrites the initial Slater determinant form of the wavefunction as $|\phi\rangle = \prod_p (a_p^\dag )^{\phi_p} |0\rangle $, representing the occupation of each molecular orbital.

The fermionic operators are used to rewrite Eq. (\ref{eq:eight}) as Eq. (\ref{eq:nine}):

\begin{eqnarray}
H = \sum_{p,q} h_{pq} a_p^\dag a_q + \frac{1}{2} \sum_{p,q,r,s} h_{pqrs} a_p^\dag a_q^\dag a_r a_s \quad .
\label{eq:nine}
\end{eqnarray}

The coefficients $h_{pq}$ and $h_{pqrs}$ encode the spatial and spin configurations of each of the electrons and depend on the inter-nuclear and inter-electron distances $R,r$. They are defined in Eq. (\ref{eq:ten}) and (\ref{eq:eleven}):

\begin{eqnarray}
h_{pq} = \int dx \phi_p^* (x) (-\frac{\nabla^2}{2} - \sum_i \frac{Z_i}{|R_i - r_i|} ) \phi_q(x) \quad ,
\label{eq:ten} 
\\
h_{pqrs} = \int dx_1 dx_2 \frac{\phi_p^*(x_1) \phi_q^*(x_2) \phi_r(x_1) \phi_s(x_2)}{|r_1 - r_2 |} \quad .
\label{eq:eleven}
\end{eqnarray}

We proceed by mapping fermionic operators, which act on Fock states of $n$ orbitals, to a Hilbert space of operators that act on spin states of $N$ qubits. This mapping corresponds to the effective interaction Hamiltonians, quantum gates and measurement basis used in quantum processors. Two useful mappings are the Jordan-Wigner transformation and the Bravyi-Kitaev transformation. The resulting Hamiltonian is expressed as a sum of tensor products involving single-qubit Pauli matrices (Eq. (\ref{eq:twelve})):

\begin{eqnarray}
H_T = \sum_{s=1}^S c_s (\otimes_{j=1}^{N} P_j^{(s)}) \quad ,
\label{eq:twelve}
\end{eqnarray}

where $P_j \epsilon \{ I,X,Y,Z \} $ , $S$ is the number of Pauli strings in the Hamiltonian and $N$ is the number of qubits. 

An example of the Hamiltonian of the $\rm H_2$ molecule for an inter-atomic distance of 1.5 \r{A} as computed by the method in hartree units is given by
\\
\\
$H_{H_2} = -0.6569 + 0.1291 \mathbf{Z_1} - 0.1291 \mathbf{Z_0} - 0.0042 \mathbf{Z_0} \mathbf{Z_1} + 0.2295 \mathbf{X_0}\mathbf{X_1}.$
\\

Examples of complete Hamiltonians found using this method for the $\rm LiH$ and $\rm BeH_2$ molecules for an inter-atomic distance of 1.5 \r{A} can be seen in Appendix~\ref{app:lih} and \ref{app:beh2}.

\section{Results}

We have coded the quantum genetic algorithm described with Python and Qadence, and we have applied it to the $\rm H_2$, $\rm LiH$ and $\rm BeH_2$ molecules. Using the Qiskit framework, combined with PySCF \cite{https://doi.org/10.1002/wcms.1340}, we have calculated the one- and two-body integrals of Eq. (\ref{eq:ten}) and (\ref{eq:eleven}), encoding the problem using the Bravyi-Kitaev method into 2 qubits for the $\rm H_2$ molecule and 6 qubits for the  $\rm LiH$ and $\rm BeH_2$ molecules. Qadence provides the high-level interface to create the quantum circuit and execute it in a Rydberg-atom simulator as described in the previous section. From Qadence (with the Pulser backend) we can also obtain the laser pulse sequences that generate the exact ansatz of interest. 

We run the algorithm for different inter-atomic distances yielding the results described in the sections below.

\subsection{$\rm H_2$ molecule}

\begin{figure}[btp]
\includegraphics[width=0.4\textwidth]{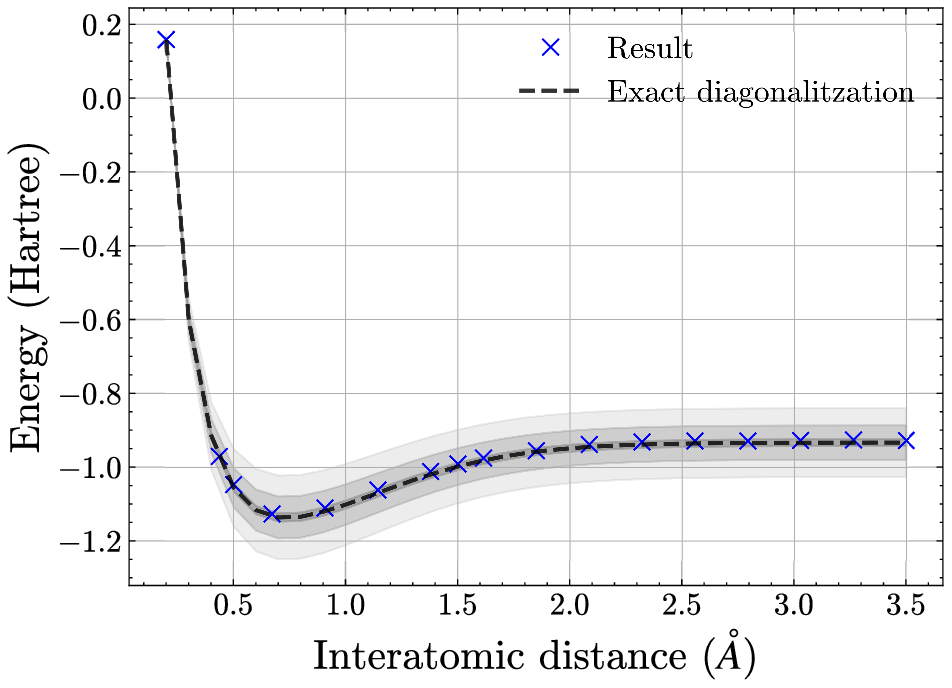}
\caption{\label{fig:h2_energies_15} Numerical results (blue marker) of the quantum genetic algorithm with our digital-analog protocol for the $\rm H_2$ molecule. The black dashed line shows the exact Hamiltonian diagonalization. The gray shades indicate respectively errors of 10, 5, and 1\%. All the points from the protocol fall within the 1\% error zone.}
\end{figure}

\begin{figure}[btp]
\includegraphics[width=0.4\textwidth]{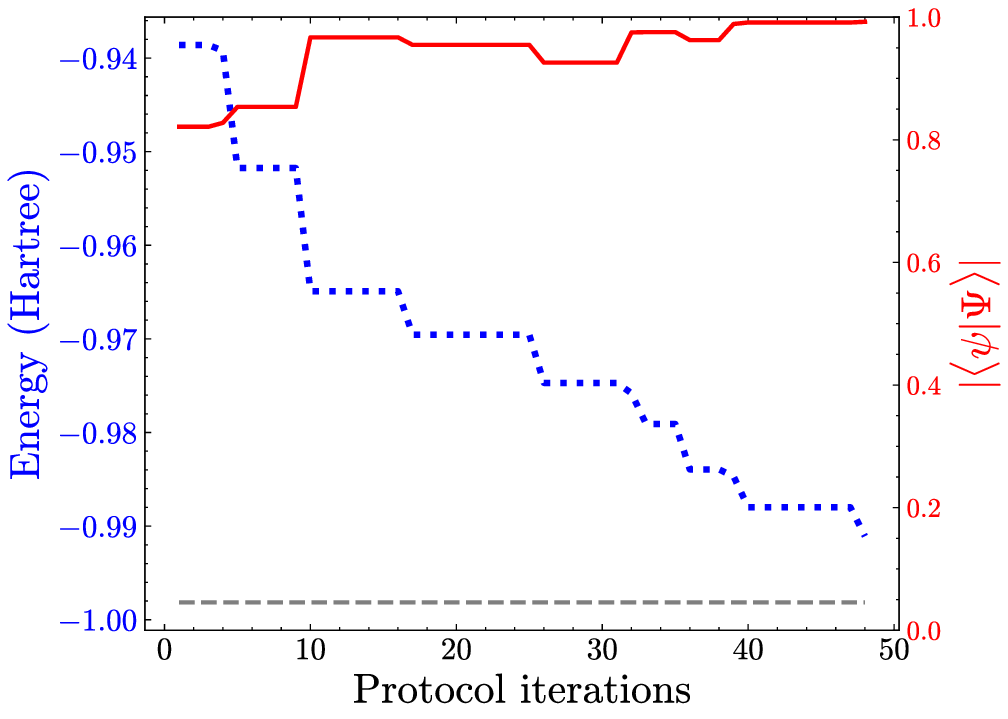}
\caption{\label{fig:h2_energy_overlap} Ground state overlap (red solid line) and energy of the best solution found (blue dotted line) as a function of the number of protocol iterations for the $\rm H_2$ molecule, for an inter-atomic distance of 1.5 \r{A}. The exact ground state energy is also shown (black dashed line).}
\vspace{-2ex}%
\end{figure}

Fig. \ref{fig:h2_energies_15} shows that the ground state energy can be obtained with an error lower than 1\% for all inter-atomic distances, ranging from 0.2 to 3.5 \r{A}. Focusing on an inter-atomic distance of 1.5 \r{A}, we show in Fig. \ref{fig:h2_energy_overlap} the ground state overlap (red line) and the energy of the best solution found (blue dotted line) as a function of the number of protocol iterations. With less than 50 iterations the state overlap was nearly 1, proving that the state obtained is a good approximation to the ground state of the molecule. In the same way, starting from a random state that has a higher energy, iteration after iteration the energy converges towards the exact ground state (black dashed line) until reaching an error lower than 1\% (our stopping criteria). The computing time for each result was less than 2 minutes. 

\begin{figure*}[btp]
\includegraphics[height=0.4\textwidth]{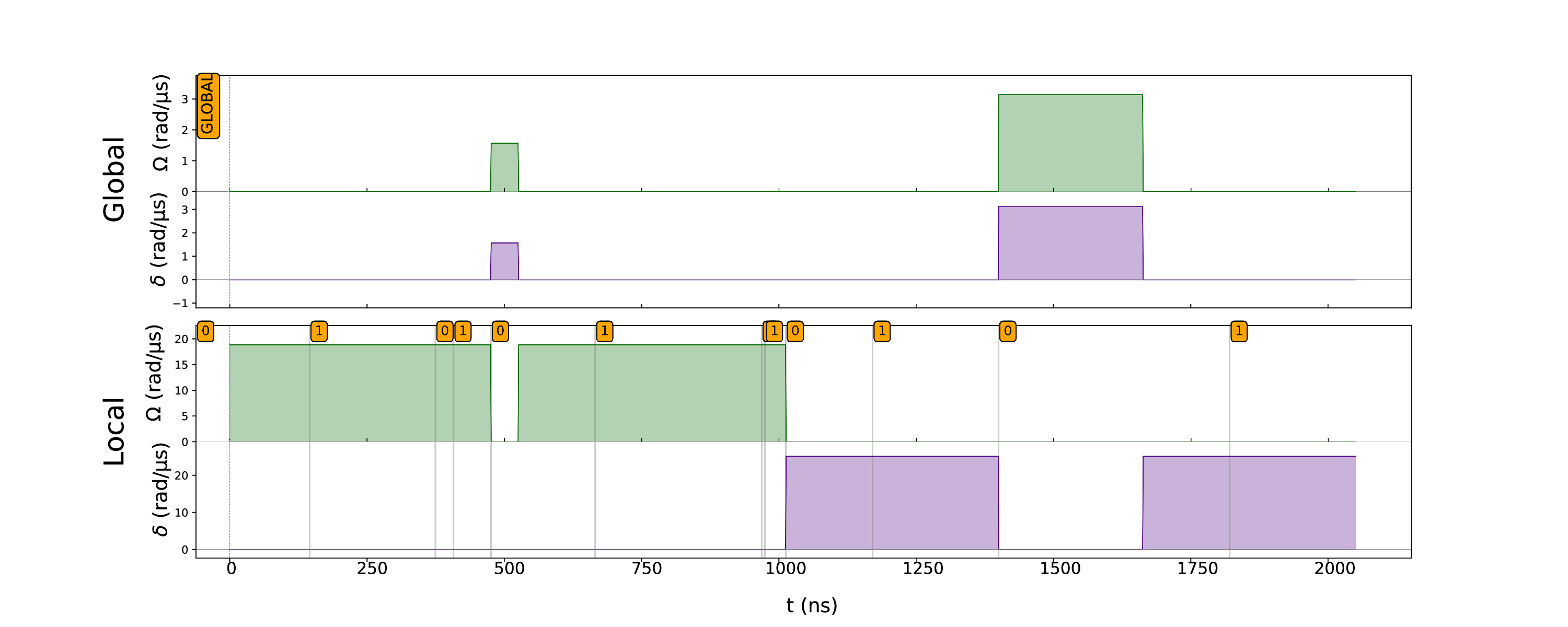}
\caption{\label{fig:H2_pulse_15} Laser pulse sequence to generate the best solution found for the $\rm H_2$ molecule and  1.5-\r{A} inter-atomic distance.}
\vspace{3ex}%
\end{figure*}

The final circuit found for the $\rm H_2$ molecule for an inter-atomic distance of 1.5 \r{A} is the one shown in Fig. \ref{fig:h2_circuit_15}, while the corresponding pulse sequence is shown in Fig. \ref{fig:H2_pulse_15}. 

The pulse sequence is divided in two regions: local and global. The local region (lower part) indicate local addressability to individual qubits, with number 0 and 1 indicating qubit 0 and qubit 1. Green and violet colors indicate Rabi ($\Omega$) and detuning ($\delta$) frequencies respectively. The horizontal axis indicates time duration in nanoseconds. The upper part (global region) indicates global addressability to all qubits at the same time. Again, green and violet indicate Rabi and detuning frequencies respectively. 

\subsection{$\rm LiH$ and $\rm BeH_2$ molecules }

\begin{figure*}[hbt!]
\centering
    \includegraphics[width=0.4\linewidth]{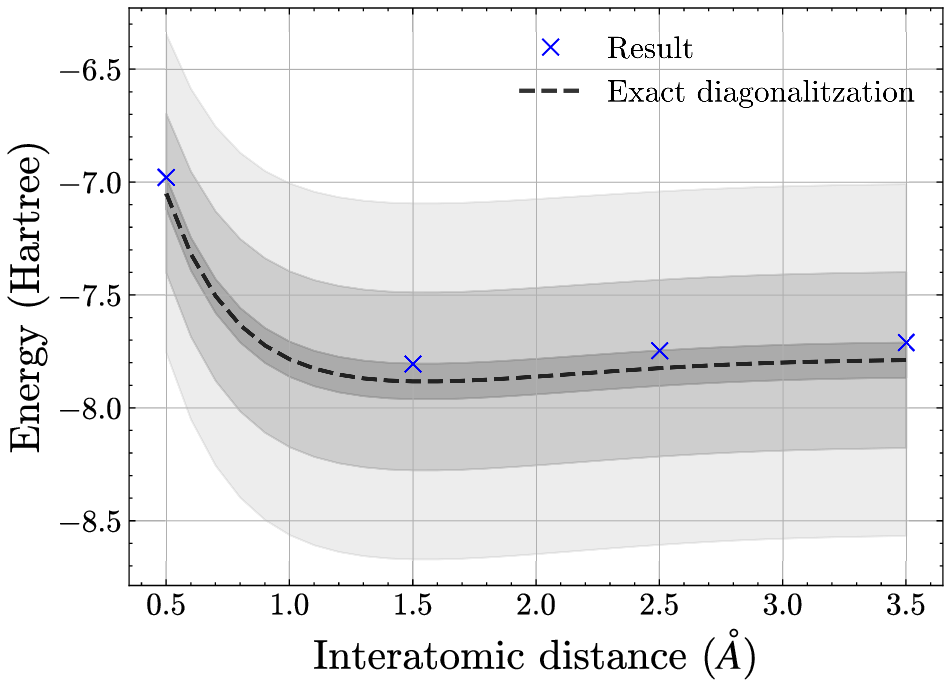}\hfil
    \includegraphics[width=0.4\linewidth]{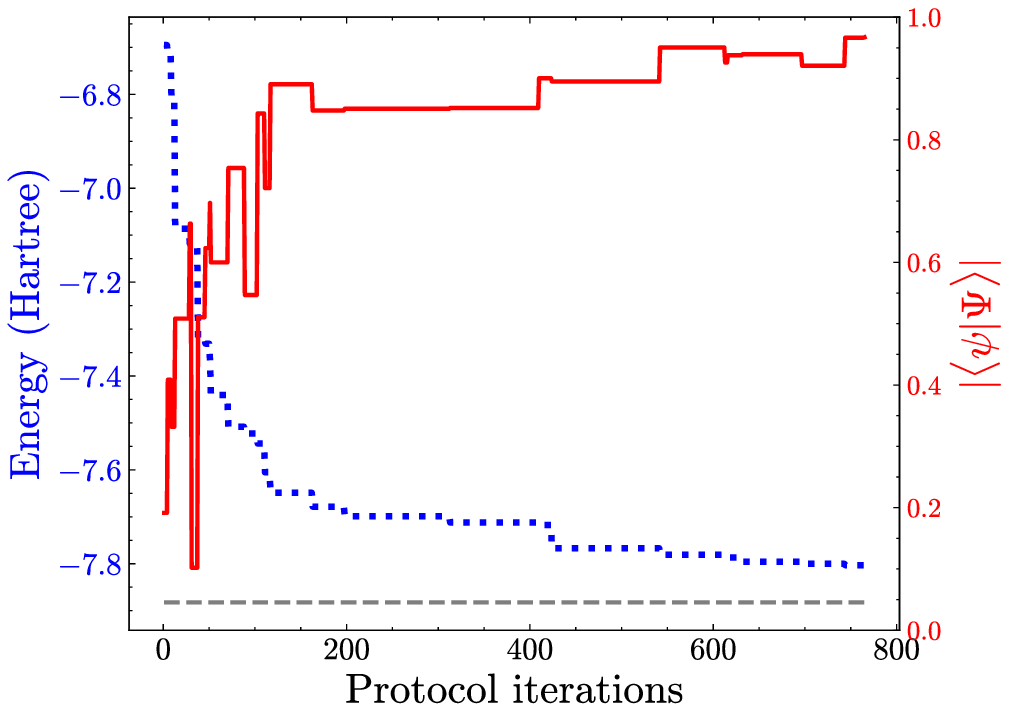}\par\medskip
    \includegraphics[width=0.4\linewidth]{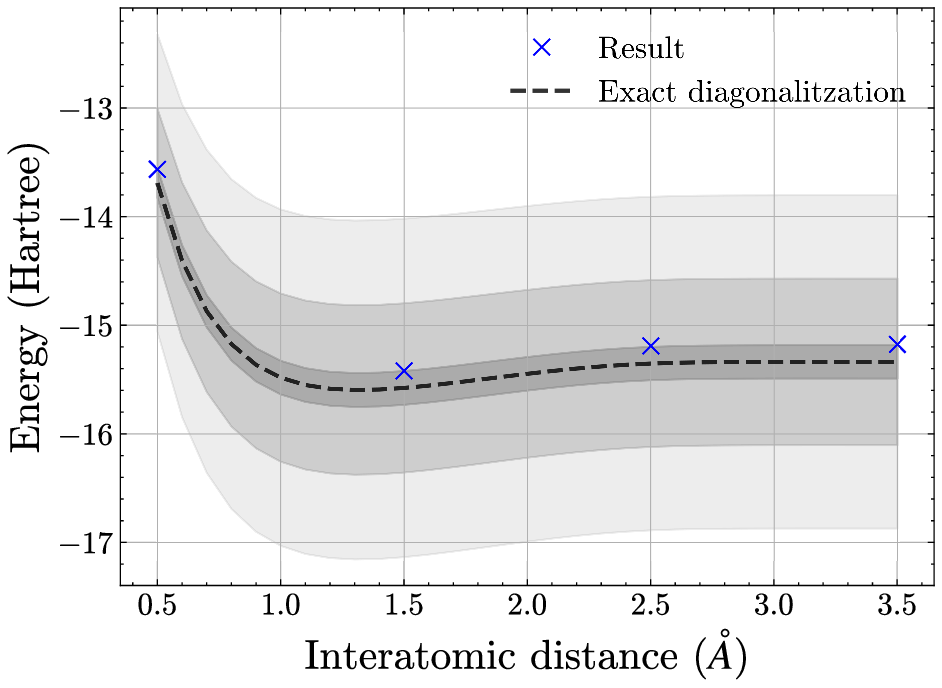}\hfil
    \includegraphics[width=0.4\linewidth]{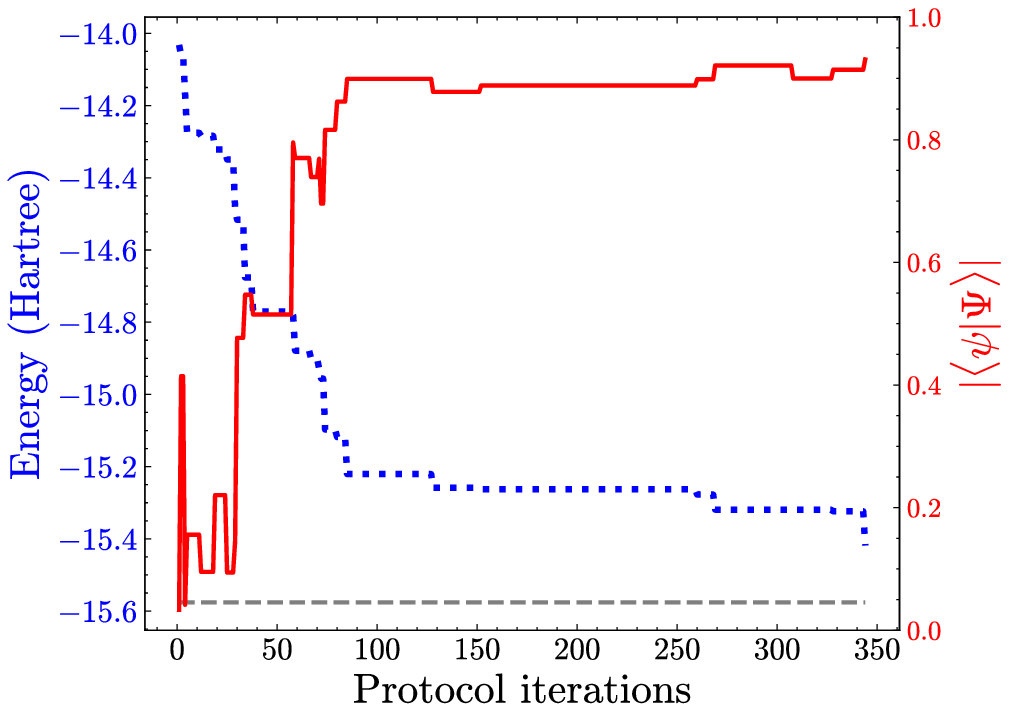}\par\medskip
\caption{\label{fig:LiH_BeH2_energies_plot}Numerical results (blue marker) of the quantum genetic algorithm with our digital-analog protocol for the $\rm LiH$ molecule (top left) and the $\rm BeH_2$ molecule (bottom left) for different inter-atomic distances. The dashed black line showns the exact Hamiltonian diagonalization and the gray shades indicate respectively errors of 10, 5, and 1\%. All the points from the protocol fall within the 1\% error zone.
For the $\rm LiH$ molecule (top right) and for the $\rm BeH_2$ molecule (bottom right) we also show the ground state overlap (red solid line) and energy of the best solution found (blue dotted line) as a function of the number of protocol iterations  for an inter-atomic distance of 1.5 \r{A}. The exact ground state energy is also shown in both cases (black dashed line).
}
\vspace{4ex}%
\end{figure*}

For the $\rm LiH$ and  $\rm BeH_2$ molecules the computing times for each result were much longer (1 to 2 days) than with the $\rm H_2$ molecule because in these two cases the 200 candidates consisted of 6 qubits each. For these two molecules we ran the algorithm for four different inter-atomic distances yielding the results shown in Fig. \ref{fig:LiH_BeH2_energies_plot} (top left for $\rm LiH$ and bottom left for $\rm BeH_2$). For both molecules, the ground state energy at different inter-atomic distances is obtained with a relative error below 1\%.

Fig. \ref{fig:LiH_BeH2_energies_plot} (top right for $\rm LiH$, bottom right for $\rm BeH_2$) shows the ground state overlap (red solid lines) for an inter-atomic distance of 1.5 \r{A}: starting from a random solution with about 0.2 state overlap, after around 800 protocol iterations for $\rm LiH$ (350 for the $\rm BeH_2$ molecule) we reach a value closer to 1 for both molecules. The same plots also show the ground state energy convergence (blue dotted line): starting from a higher energy, for iteration after iteration the protocol succeeds to minimize the energy until reaching an error below 1\% for the two molecules.

\begin{figure*}[btp]
\includegraphics[width=\textwidth]{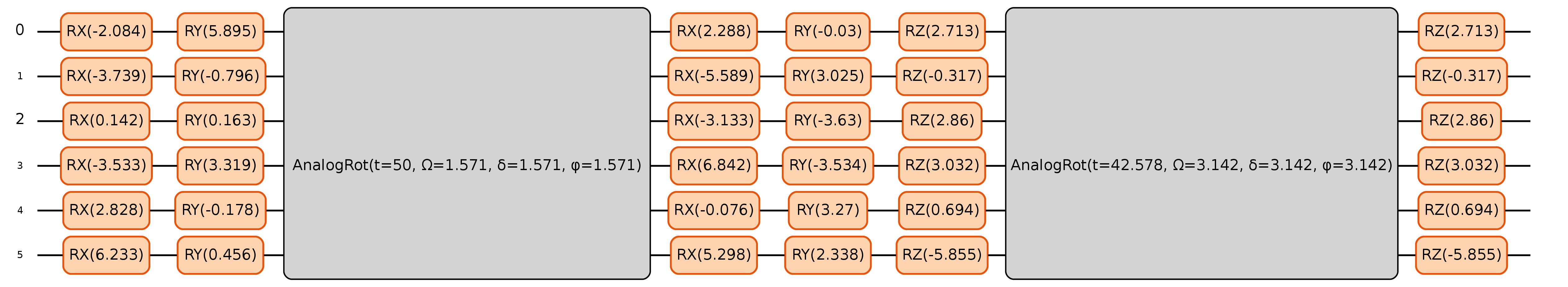}
\caption{\label{fig:LiH_circuit} Best six qubit circuit found using the protocol for the $\rm LiH$ molecule with an inter-atomic distance of 1.5 \r{A}.}
\end{figure*}

\begin{figure*}[btp]
\includegraphics[height=0.4\textwidth]{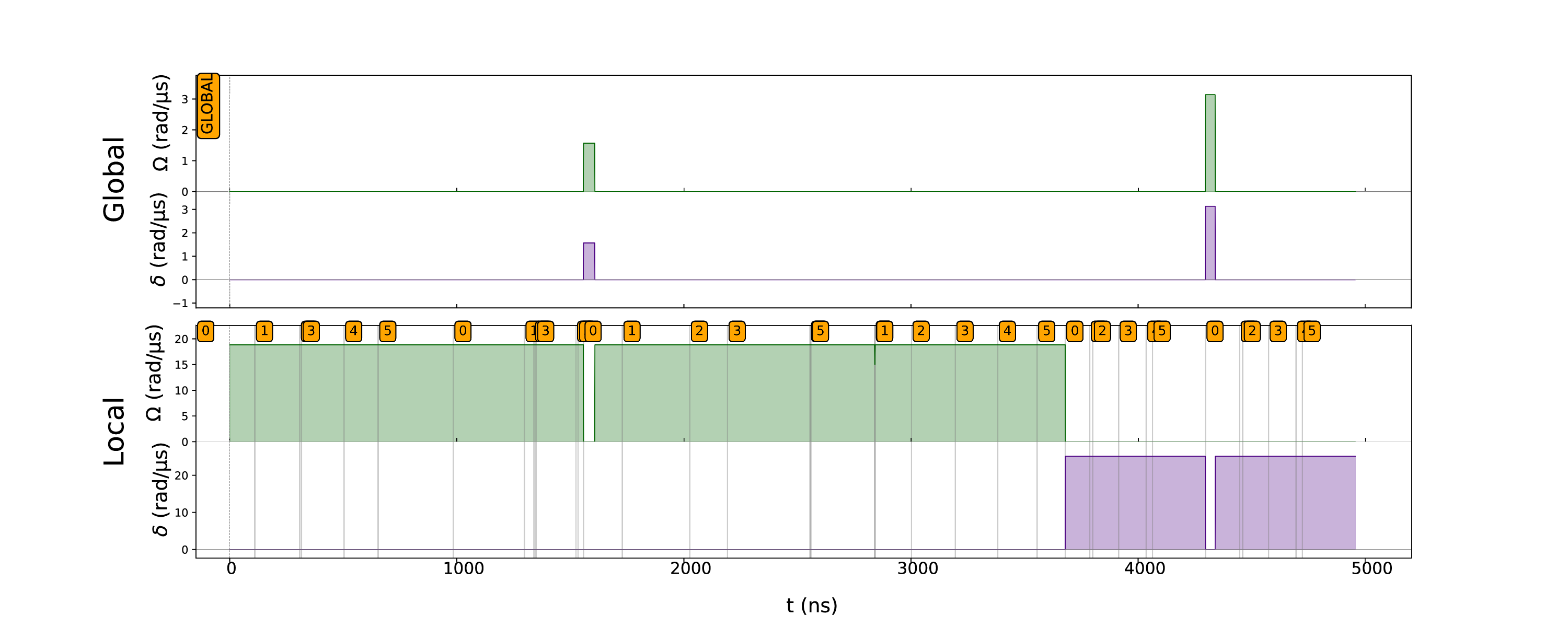}
\caption{\label{fig:LiH_pulse} Laser pulse sequence to generate the best solution found for the $\rm LiH$ molecule for an inter-atomic distance of 1.5 \r{A}.}
\end{figure*}

The corresponding circuit for the $\rm LiH$ molecule for a distance of 1.5 \r{A} can be seen in Fig. \ref{fig:LiH_circuit}. We can see that each qubit has different rotations in RX, RY, and RZ, and the first global analog block has a 50 ns time as defined initially while the second global analog block has a 42 ns evolution time (decided by the protocol). The pulse sequence diagram can also be seen in Fig. \ref{fig:LiH_pulse}.

\begin{figure*}[btp]
\includegraphics[width=\textwidth]{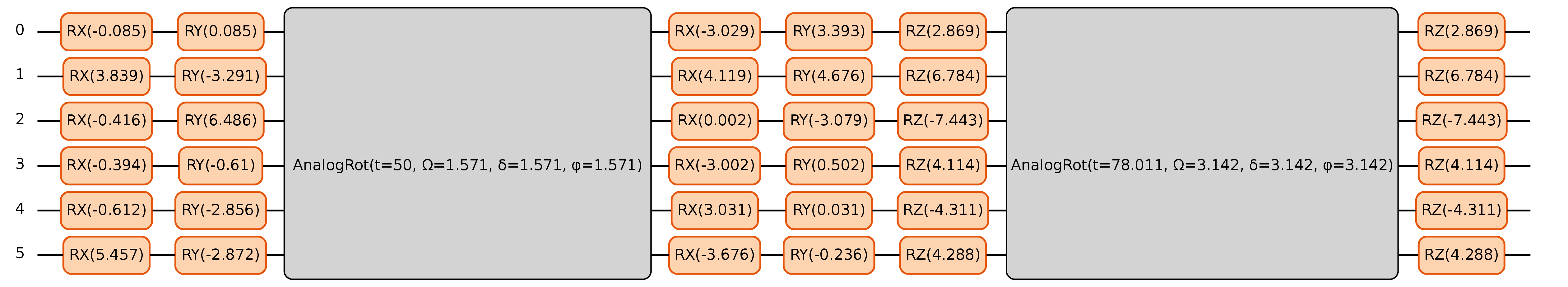}
\caption{\label{fig:BeH2_circuit} Best six qubit circuit found using the protocol for the $\rm BeH_2$ molecule with an inter-atomic distance of 1.5 \r{A}.}
\vspace{-1.2ex}%
\end{figure*}

\begin{figure*}[btp]
\vspace{-1.1ex}%
\includegraphics[height=0.4\textwidth]{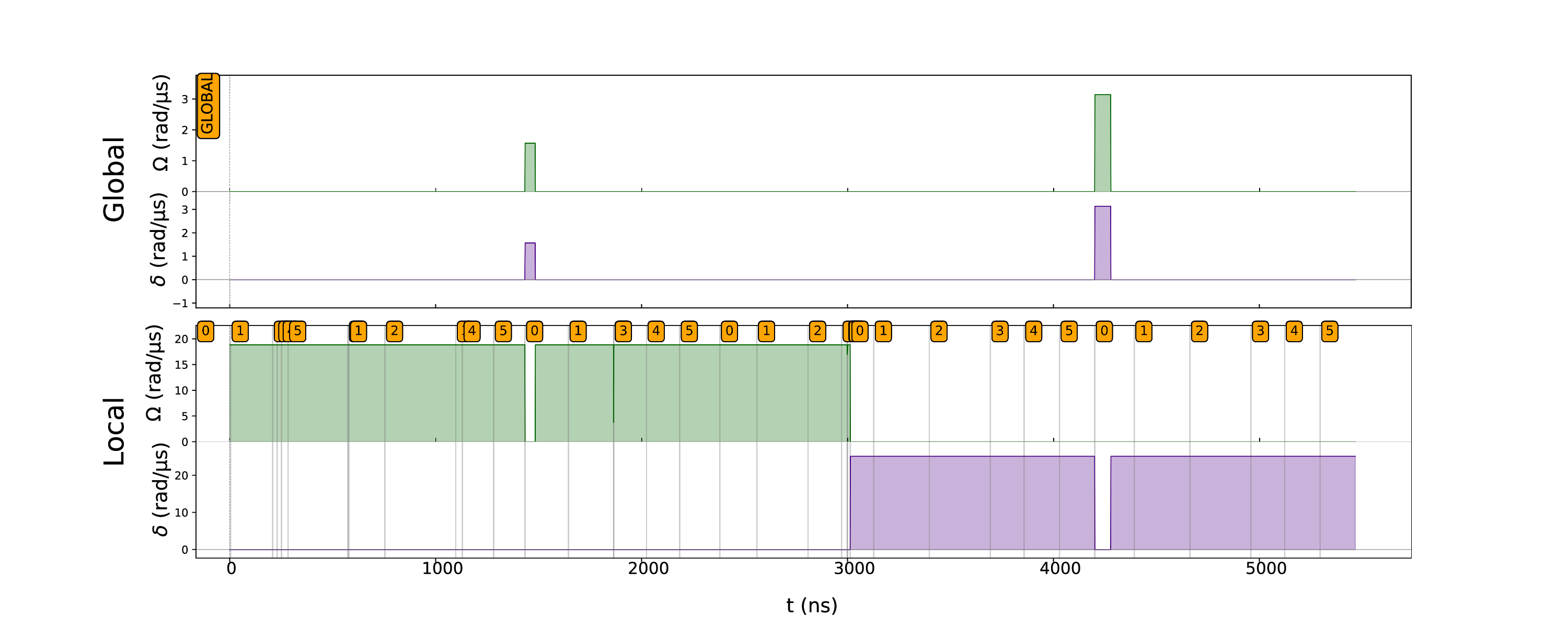}
\caption{\label{fig:BeH2_pulse} Laser pulse sequence to generate the best solution found for the $\rm BeH_2$ molecule for an inter-atomic distance of 1.5 \r{A}.}
\end{figure*}

The corresponding circuit for the $\rm BeH_2$ molecule for a distance of 1.5 Å can be seen in Fig. \ref{fig:BeH2_circuit}. The second global analog block has a 78 ns time. The pulse sequence diagram can also be seen in Fig. \ref{fig:BeH2_pulse}.

\section{Discussion}

In this paper, we have proposed a quantum genetic algorithm based on the DAQC paradigm and we have tested it with Hamiltonians coming from chemistry using a backend that emulates Rydberg quantum processors. A quantum genetic algorithm is proposed under the DAQC realm, and our purpose was to prove that protocols like the ones described in this paper are feasible and give positive outcomes. Our numerical results should be viewed as a testing benchmark and should trigger further explorations.  

\begin{figure}
\includegraphics[width=0.3\textwidth]{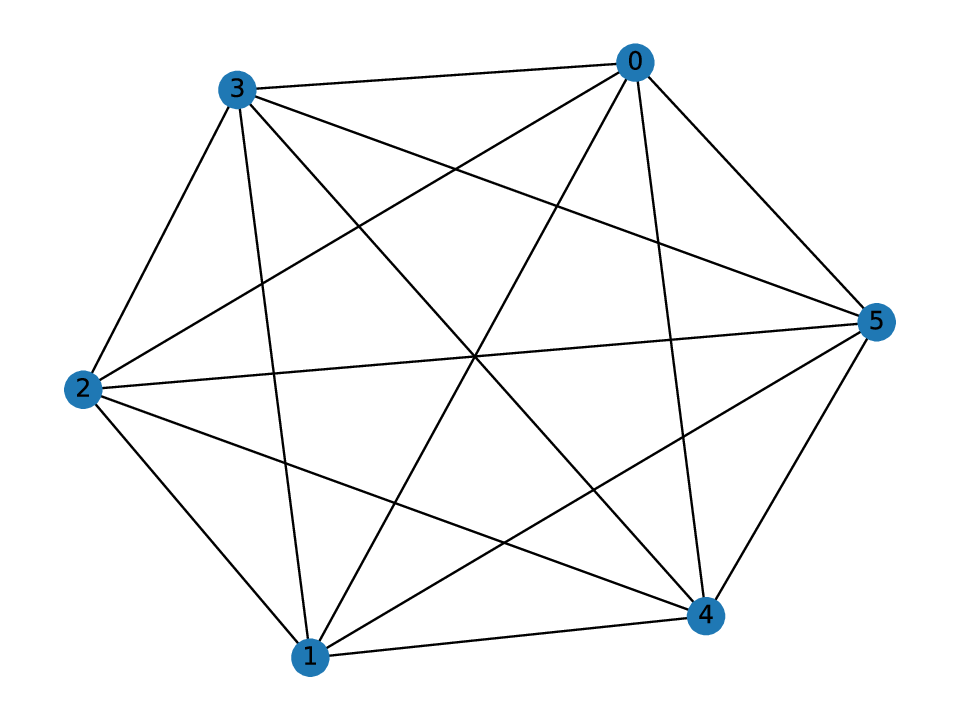}
\caption{\label{fig:topology} Topology of the register of the 6 qubits with all-to-all connectivity.}
\vspace{-3ex}%
\end{figure}

The choice of ansatz that was used through all the experiments was not specially selected for these particular Hamiltonians. In contrast, we selected an ansatz that could provide enough diversity and entanglement and we were restricted to the operations that Qadence (working with the Pulser backend) allowed us to use: single qubit rotations and a global driving Hamiltonian according to the Rydberg interaction. 
In addition, the Rydberg-atom–atom distance as well as the register topologies can play a significant role in modulating the convergence of protocols like this one. In this paper we have not used this as a variable but it could be also incorporated into the quantum genetic algorithm protocol in further experiments. For the whole experimentation the atom-atom distance was set at $8 \mu m$ (typical distance for standard experimental setup of neutral atoms) and the interaction was set to all-to-all (see Fig. \ref{fig:topology}).

All in all, the aim of the paper was to keep the quantum genetic algorithm as simple as possible, with just a small set of variational parameters, and give evidence that it converges towards the desired solution. Our quantum genetic protocol employs 200 candidate solutions (circuits) at the same time per iteration. This inherent parallelism allows us to explore multiple solutions simultaneously, potentially accelerating the optimization. Additionally, genetic algorithms exhibit remarkable adaptability, enabling us to explore various pathways concurrently and avoid getting trapped in local solutions.

In addition, digital-analog can benefit from the fact that analog operations can be performed with much higher fidelities than when using digital gates. We numerically estimate the gate fidelity (Eq. (\ref{eq:fidelity})) by discretizing the integral $d \tilde{U}$ over 500 samples of the noise model following the same method as described in \cite{lu2024digital}: \\
\\
\vfill
\begin{eqnarray}
F(U) = \int d \tilde{U} \int d\Psi \braket{\Psi | U^\dag \tilde{U} | \Psi} \quad .
\label{eq:fidelity}
\end{eqnarray}

We can ignore all single-qubit gates as they generally have much higher fidelity than multiqubit gates. For simplicity, we only consider a model of coherent errors independent among every time evolution operation that captures noise in the Rabi frequencies and detunings as well as position perturbations of each atom. A typical range of experimental values can be extracted from the literature \cite{lu2024digital}: (1) noisy Rabi frequency,  $\tilde{\Omega} = \Omega \cdot N(1, 0.01) $; (2) noisy detuning, $\tilde{\delta} = \delta + N(0, 0.6 rad / \mu s)$; and (3) perturbed atomic coordinates according to a Gaussian process, $\tilde{r_j} = r_j + N(0, 0.1 \mu m)$, where $N(\mu, \sigma)$ is a Gaussian distribution with mean $\mu$ and variance $\sigma^2$.

\begin{figure}
\includegraphics[width=0.8\linewidth]{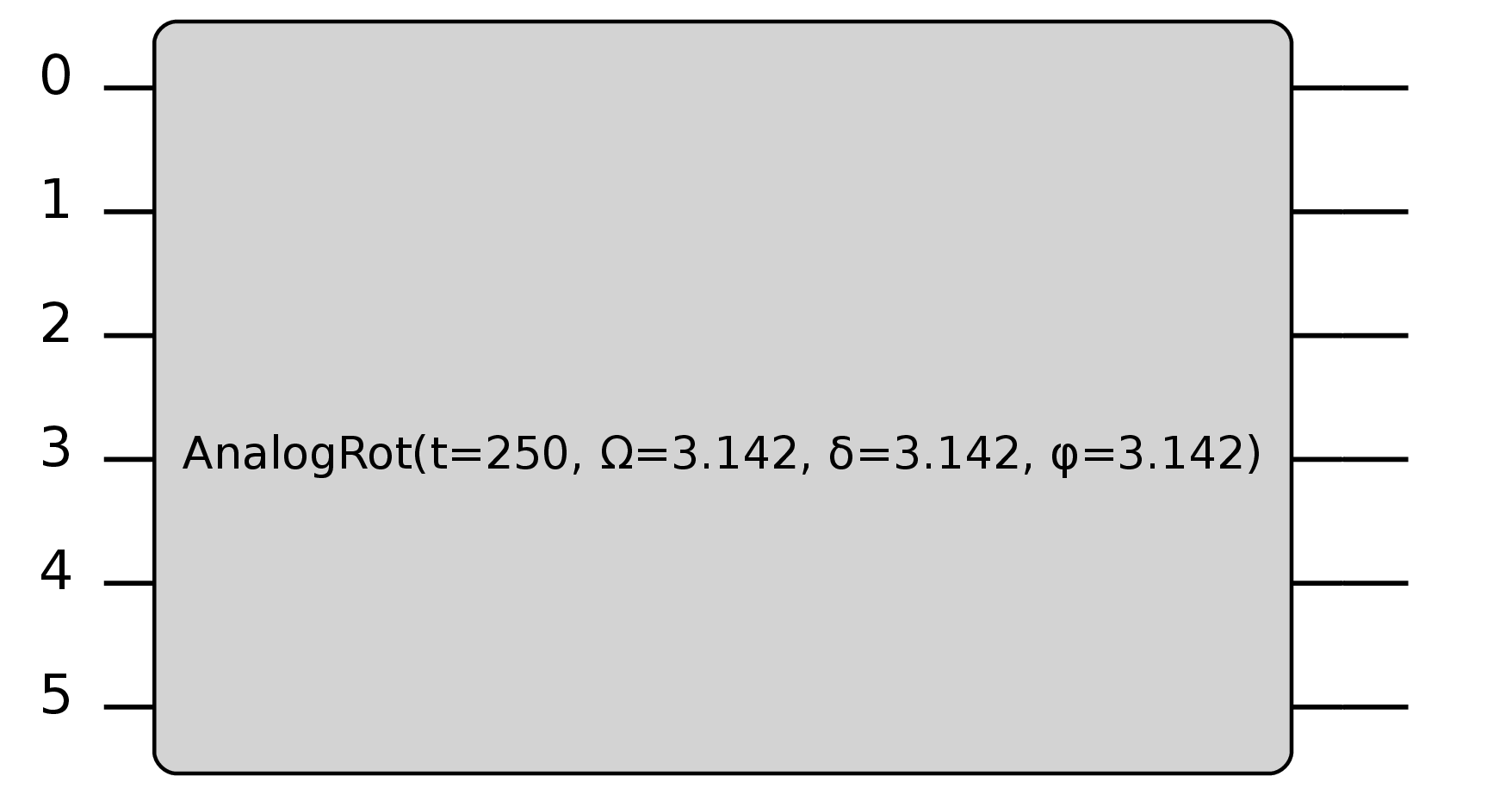}
\vspace{4.00mm}
\includegraphics[width=0.8\linewidth]{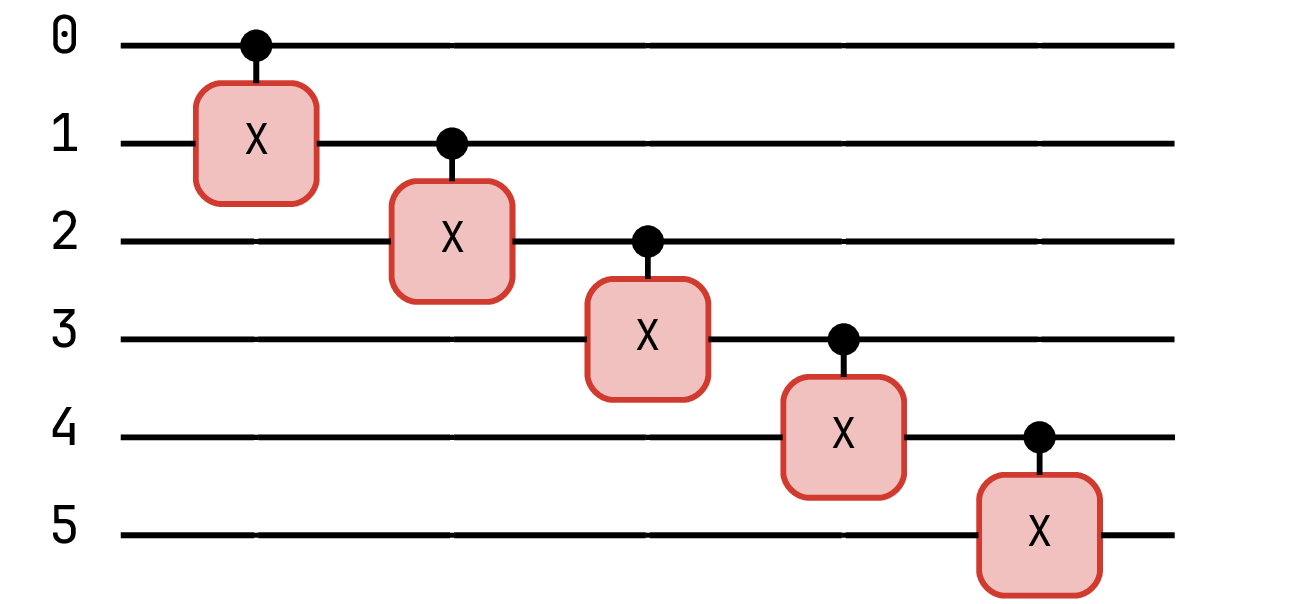}
 \caption{\label{fig:analog_cnot}%
Circuits to generate entanglement and diversity in the solution space: (top) Analog: global driving Rydberg Hamiltonian as the one used in the protocol; (bottom) Fully digital circuit with CNOTs.
 }
\end{figure}

For the sake of providing diversity and entanglement between qubits, an equivalent fully digital quantum circuit could be constructed with a set of CNOT gates connecting all qubits. This would be equivalent in terms of achieving similar functionality (entanglement) although the circuits and the final unitary they generate are not the same. For the case of a 6 qubit ansatz, Fig. \ref{fig:analog_cnot} shows an example.

Noise for CNOT gates will depend on the physical system used to implement it. In this estimation, we have modelled the CNOT as two Hadamards with a CPHASE gate with $\phi = \pi$ including the noise in the CPHASE angle. As a simple but realistic digital independent noise model, we sample $\phi$ from a Gaussian distribution with mean $\mu = \pi$ and standard deviation $\sigma$ chosen such that the resulting gate fidelity is 99\%. 

\begin{figure}
\includegraphics[width=0.8\linewidth]{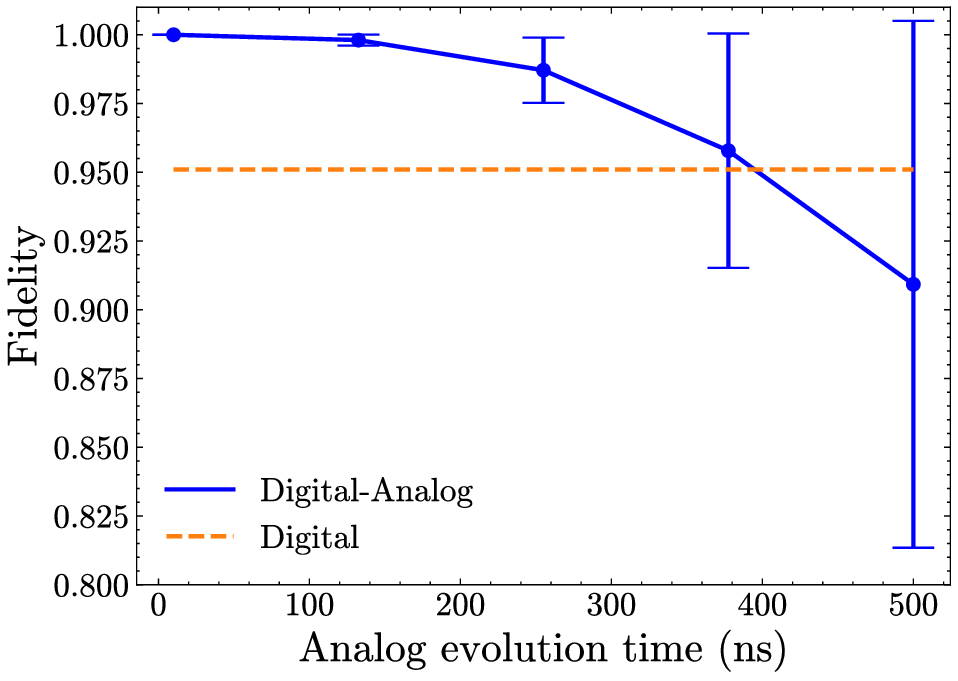}
\vspace{1.00mm}
\vspace{1.00mm}
\includegraphics[width=0.8\linewidth]{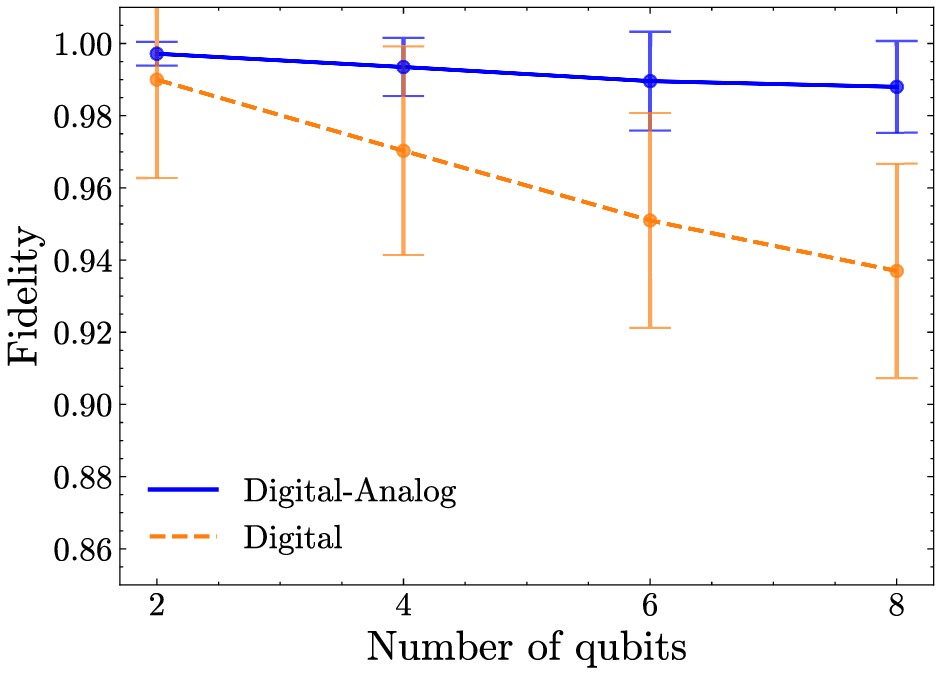}
 \caption{\label{fig:fidelity}%
(top) Estimated fidelity of the analog block of 6 qubits by varying the applied evolution time and compared with the fidelity of the equivalent digital CNOT circuit for also 6 qubits. (bottom) Estimated gate fidelity for the analog block with a fixed evolution time of 250 ns, compared with the fidelity of the equivalent digital CNOT circuit as a function of the number of qubits.
 }
\end{figure}

For 6 qubits, we show in Fig. \ref{fig:fidelity} (top) the analog block fidelity as a function of the evolution time, compared with the fidelity of an equivalent digital block made of 5 CNOTs (such as in  Fig. \ref{fig:analog_cnot}). We can see that when the evolution time is lower than approximately 400 ns the digital-analog block has a higher gate fidelity than the fully digital circuit.

We can also explore how the gate fidelity would change by changing the number of qubits (Fig. \ref{fig:fidelity} (bottom)). For this we set a fixed evolution time for 250 ns for the analog block and we set the equivalent number of CNOTs as one less than the number of qubits. The graph shows that the separation between the digital and digital-analog lines increases with the number of qubits, suggesting that for a large number of qubits the digital-analog protocol would perform at much higher fidelities than an equivalent digital one. Extending the comparison to more qubits would offer deeper insights, but classical simulators currently struggle with the increased computational load. Future work could explore this scaling behavior using real quantum devices.

In this paper, we employed a simplified noise model that considers coherent errors arising from noise in the Rabi frequencies, detunings, and position perturbations of each atom. We did not include the effects of laser phase noise \cite{de2018analysis, lee2019coherent} or other complex noise sources in our analysis. Our conclusions are drawn from this simplified noise model, and we emphasize the need for further experimental validation on real Rydberg-atom systems to determine the robustness of AQC against more realistic noise sources, including laser phase noise.

\section{Conclusions}

In this paper we presented a hybrid quantum genetic algorithm developed under the DAQC framework and simulated it in a Rydberg-atom emulator. The performance of the algorithm was evaluated by finding the ground state of Hamiltonians coming from chemistry: the $\rm H_2$, $\rm LiH$, and $\rm BeH_2$ molecules. 

The results show that the algorithm is able to find the ground state energy of such molecules with less than 1\% error and a state overlap nearing 1. In the case of the $\rm H_2$ molecule (2 qubit circuits) the computing times are negligible, while in the case of $\rm LiH$ and $\rm BeH_2$ molecules (6 qubits circuits) the running times were between one and two days. 

The higher gate fidelities of global analog operations compared to an equivalent purely digital circuit indicate that the digital-analog paradigm is a promising quantum computing option within the NISQ era. 

\vspace{-1ex}%
\begin{acknowledgments}
\vspace{-2ex}%
This work has been financially supported by the Ministry for Digital Transformation and Civil Service of the Spanish Government through the QUANTUM ENIA project Quantum Spain, and by “European Union NextGenerationEU/PRTR” within the framework of the “Digital Spain 2026 Agenda”. 
\end{acknowledgments}
\pagebreak[4]

\begin{widetext}

\appendix*

\section{Hamiltonians for $\rm LiH$ and $\rm BeH_2$ molecules}

In this section, we show examples of complete Hamiltonians for the $\rm LiH$ and $\rm BeH_2$ molecules obtained with the method described in the paper, in both cases for an inter-atomic distance of 1.5 \r{A}.  

\subsection{\label{app:lih} $\rm LiH$ molecule}

    $H_{LiH} = -5.5546 -0.3010\mathbf{Z_5} + 0.0049\mathbf{Z_0}\mathbf{Z_1}\mathbf{Z_2}\mathbf{Z_3}\mathbf{X_5} -0.0049\mathbf{X_5} + 0.0059\mathbf{X_3}\mathbf{X_4}\mathbf{X_5} + 0.0059\mathbf{X_3}\mathbf{Y_4}\mathbf{Y_5} -0.3899\mathbf{Z_0}\mathbf{Z_1}\mathbf{Z_2}\mathbf{Z_3}\mathbf{Z_5} -0.0236\mathbf{X_3}\mathbf{X_4}\mathbf{Z_5} -0.0236\mathbf{Z_0}\mathbf{Z_1}\mathbf{Z_2}\mathbf{Y_3}\mathbf{Y_4} + 0.8490\mathbf{Z_0}\mathbf{Z_2}\mathbf{Z_3}\mathbf{Z_4} -0.8490\mathbf{Z_1} -0.5773\mathbf{Z_4} -0.3010\mathbf{Z_3}\mathbf{Z_4} -0.0049\mathbf{Y_2}\mathbf{Y_3}\mathbf{Z_4} -0.0049\mathbf{X_2}\mathbf{X_3} -0.0059\mathbf{X_0}\mathbf{X_3} -0.0059\mathbf{Y_0}\mathbf{Z_1}\mathbf{Y_3}\mathbf{Z_4} -0.3899\mathbf{Z_2} + 0.0236\mathbf{X_0}\mathbf{X_2}\mathbf{Z_3}\mathbf{Z_4} + 0.0236\mathbf{Y_0}\mathbf{Z_1}\mathbf{Y_2}\mathbf{Z_3}\mathbf{Z_4} -0.5773\mathbf{Z_0}\mathbf{Z_1} + 0.0536\mathbf{Z_0}\mathbf{Z_1}\mathbf{Z_2}\mathbf{Z_3} -0.0044\mathbf{X_3}\mathbf{X_4} -0.0044\mathbf{Z_0}\mathbf{Z_1}\mathbf{Z_2}\mathbf{Y_3}\mathbf{Y_4}\mathbf{Z_5} -0.1316\mathbf{Z_0}\mathbf{Z_2}\mathbf{Z_3}\mathbf{Z_4}\mathbf{Z_5} + 0.1316\mathbf{Z_1}\mathbf{Z_5} + 0.0836\mathbf{Z_4}\mathbf{Z_5} + 0.1236\mathbf{Z_3}\mathbf{Z_4}\mathbf{Z_5} + 0.0117\mathbf{Y_2}\mathbf{Y_3}\mathbf{Z_4}\mathbf{Z_5} + 0.0117\mathbf{X_2}\mathbf{X_3}\mathbf{Z_5} -0.0331\mathbf{X_0}\mathbf{X_3}\mathbf{Z_5} -0.0331\mathbf{Y_0}\mathbf{Z_1}\mathbf{Y_3}\mathbf{Z_4}\mathbf{Z_5} + 0.0567\mathbf{Z_2}\mathbf{Z_5} + 0.0127\mathbf{X_0}\mathbf{X_2}\mathbf{Z_3}\mathbf{Z_4}\mathbf{Z_5} + 0.0127\mathbf{Y_0}\mathbf{Z_1}\mathbf{Y_2}\mathbf{Z_3}\mathbf{Z_4}\mathbf{Z_5} + 0.1143\mathbf{Z_0}\mathbf{Z_1}\mathbf{Z_5} + 0.0045\mathbf{Z_0}\mathbf{Z_1}\mathbf{Z_2}\mathbf{Y_3}\mathbf{X_4}\mathbf{Y_5} -0.0045\mathbf{Z_0}\mathbf{Z_1}\mathbf{Z_2}\mathbf{Y_3}\mathbf{Y_4}\mathbf{X_5} + 0.0024\mathbf{Z_1}\mathbf{Z_4}\mathbf{X_5} -0.0024\mathbf{Z_0}\mathbf{Z_2}\mathbf{Z_3}\mathbf{Z_4}\mathbf{X_5} -0.0024\mathbf{Z_0}\mathbf{Z_2}\mathbf{Z_3}\mathbf{X_5} + 0.0024\mathbf{Z_1}\mathbf{X_5} -0.0028\mathbf{Z_0}\mathbf{Z_1}\mathbf{Z_2}\mathbf{Z_3}\mathbf{Z_4}\mathbf{X_5} + 0.0028\mathbf{Z_4}\mathbf{X_5} -0.0117\mathbf{Z_0}\mathbf{Z_1}\mathbf{Z_2}\mathbf{Z_4}\mathbf{X_5} + 0.0117\mathbf{Z_3}\mathbf{Z_4}\mathbf{X_5} + 0.0030\mathbf{Z_0}\mathbf{Z_1}\mathbf{X_2}\mathbf{X_3}\mathbf{Z_4}\mathbf{X_5} + 0.0030\mathbf{Z_0}\mathbf{Z_1}\mathbf{Y_2}\mathbf{Y_3}\mathbf{X_5} + 0.0030\mathbf{Y_2}\mathbf{Y_3}\mathbf{Z_4}\mathbf{X_5} + 0.0030\mathbf{X_2}\mathbf{X_3}\mathbf{X_5} -0.0084\mathbf{Y_0}\mathbf{Z_1}\mathbf{Z_2}\mathbf{Y_3}\mathbf{X_5} -0.0084\mathbf{X_0}\mathbf{Z_2}\mathbf{X_3}\mathbf{Z_4}\mathbf{X_5} -0.0084\mathbf{X_0}\mathbf{X_3}\mathbf{X_5} -0.0084\mathbf{Y_0}\mathbf{Z_1}\mathbf{Y_3}\mathbf{Z_4}\mathbf{X_5} + 0.0015\mathbf{Z_0}\mathbf{Z_1}\mathbf{Z_3}\mathbf{X_5} -0.0015\mathbf{Z_2}\mathbf{X_5} + 0.0021\mathbf{Y_0}\mathbf{Z_1}\mathbf{Y_2}\mathbf{Z_4}\mathbf{X_5} + 0.0021\mathbf{X_0}\mathbf{X_2}\mathbf{Z_4}\mathbf{X_5} + 0.0021\mathbf{X_0}\mathbf{X_2}\mathbf{Z_3}\mathbf{Z_4}\mathbf{X_5} + 0.0021\mathbf{Y_0}\mathbf{Z_1}\mathbf{Y_2}\mathbf{Z_3}\mathbf{Z_4}\mathbf{X_5} -0.0105\mathbf{Z_2}\mathbf{Z_3}\mathbf{X_5} + 0.0105\mathbf{Z_0}\mathbf{Z_1}\mathbf{X_5} -0.0006\mathbf{Z_0}\mathbf{Z_2}\mathbf{Y_3}\mathbf{Y_4}\mathbf{X_5} + 0.0006\mathbf{Z_0}\mathbf{Z_2}\mathbf{Y_3}\mathbf{X_4}\mathbf{Y_5} -0.0060\mathbf{Z_0}\mathbf{Z_1}\mathbf{X_3}\mathbf{Z_4}\mathbf{X_5} -0.0060\mathbf{Z_1}\mathbf{Z_2}\mathbf{X_3}\mathbf{Z_4}\mathbf{X_5} + 0.0060\mathbf{Z_1}\mathbf{Z_2}\mathbf{Y_3}\mathbf{Y_5} + 0.0060\mathbf{Z_0}\mathbf{Z_1}\mathbf{Y_3}\mathbf{Y_5} -0.0048\mathbf{Z_0}\mathbf{Z_1}\mathbf{X_2}\mathbf{Z_3}\mathbf{X_5} -0.0048\mathbf{Z_1}\mathbf{X_2}\mathbf{Z_4}\mathbf{X_5} + 0.0048\mathbf{Z_1}\mathbf{Y_2}\mathbf{Z_4}\mathbf{Y_5} + 0.0048\mathbf{Z_0}\mathbf{Z_1}\mathbf{Y_2}\mathbf{Z_3}\mathbf{Y_5} + 0.0049\mathbf{X_0}\mathbf{Z_1}\mathbf{Z_2}\mathbf{Z_3}\mathbf{X_5} + 0.0049\mathbf{X_0}\mathbf{Z_4}\mathbf{X_5} -0.0049\mathbf{Y_0}\mathbf{Z_1}\mathbf{Z_4}\mathbf{Y_5} -0.0049\mathbf{Y_0}\mathbf{Z_2}\mathbf{Z_3}\mathbf{Y_5} -0.0006\mathbf{Z_1}\mathbf{X_3}\mathbf{X_4}\mathbf{X_5} -0.0006\mathbf{Z_1}\mathbf{X_3}\mathbf{Y_4}\mathbf{Y_5} -0.0060\mathbf{Y_0}\mathbf{Y_1}\mathbf{X_3}\mathbf{Z_4}\mathbf{X_5} + 0.0060\mathbf{Y_0}\mathbf{X_1}\mathbf{Y_3}\mathbf{X_5} + 0.0060\mathbf{Y_0}\mathbf{X_1}\mathbf{X_3}\mathbf{Z_4}\mathbf{Y_5} + 0.0060\mathbf{Y_0}\mathbf{Y_1}\mathbf{Y_3}\mathbf{Y_5} -0.0048\mathbf{Y_0}\mathbf{Y_1}\mathbf{X_2}\mathbf{Z_3}\mathbf{X_5} + 0.0048\mathbf{Y_0}\mathbf{X_1}\mathbf{Y_2}\mathbf{Z_3}\mathbf{X_5} + 0.0048\mathbf{Y_0}\mathbf{X_1}\mathbf{X_2}\mathbf{Z_3}\mathbf{Y_5} + 0.0048\mathbf{Y_0}\mathbf{Y_1}\mathbf{Y_2}\mathbf{Z_3}\mathbf{Y_5} + 0.0049\mathbf{Z_0}\mathbf{X_1}\mathbf{Z_4}\mathbf{X_5} -0.0049\mathbf{X_1}\mathbf{Z_4}\mathbf{X_5} + 0.0049\mathbf{Z_0}\mathbf{Y_1}\mathbf{Z_4}\mathbf{Y_5} -0.0049\mathbf{Y_1}\mathbf{Z_4}\mathbf{Y_5} -0.0331\mathbf{Y_3}\mathbf{Y_4}\mathbf{X_5} + 0.0331\mathbf{Y_3}\mathbf{X_4}\mathbf{Y_5} + 0.0084\mathbf{Y_2}\mathbf{Z_3}\mathbf{Y_4}\mathbf{X_5} + 0.0084\mathbf{X_2}\mathbf{X_4}\mathbf{X_5} -0.0084\mathbf{Y_2}\mathbf{Z_3}\mathbf{X_4}\mathbf{Y_5} + 0.0084\mathbf{X_2}\mathbf{Y_4}\mathbf{Y_5} -0.0307\mathbf{X_0}\mathbf{X_4}\mathbf{X_5} -0.0307\mathbf{Y_0}\mathbf{Z_1}\mathbf{Z_3}\mathbf{Y_4}\mathbf{X_5} -0.0307\mathbf{X_0}\mathbf{Y_4}\mathbf{Y_5} + 0.0307\mathbf{Y_0}\mathbf{Z_1}\mathbf{Z_3}\mathbf{X_4}\mathbf{Y_5} -0.0024\mathbf{Z_2}\mathbf{X_3}\mathbf{X_4}\mathbf{X_5} -0.0024\mathbf{Z_2}\mathbf{X_3}\mathbf{Y_4}\mathbf{Y_5} -0.0078\mathbf{X_0}\mathbf{X_2}\mathbf{Y_3}\mathbf{Y_4}\mathbf{X_5} -0.0078\mathbf{Y_0}\mathbf{Z_1}\mathbf{Y_2}\mathbf{Y_3}\mathbf{Y_4}\mathbf{X_5} + 0.0078\mathbf{X_0}\mathbf{X_2}\mathbf{Y_3}\mathbf{X_4}\mathbf{Y_5} + 0.0078\mathbf{Y_0}\mathbf{Z_1}\mathbf{Y_2}\mathbf{Y_3}\mathbf{X_4}\mathbf{Y_5} + 0.0351\mathbf{Z_0}\mathbf{Z_1}\mathbf{X_3}\mathbf{X_4}\mathbf{X_5} + 0.0351\mathbf{Z_0}\mathbf{Z_1}\mathbf{X_3}\mathbf{Y_4}\mathbf{Y_5} -0.1308\mathbf{Z_1}\mathbf{Z_4}\mathbf{Z_5} + 0.1308\mathbf{Z_0}\mathbf{Z_2}\mathbf{Z_3}\mathbf{Z_5} + 0.0539\mathbf{Z_0}\mathbf{Z_1}\mathbf{Z_2}\mathbf{Z_3}\mathbf{Z_4}\mathbf{Z_5} + 0.0567\mathbf{Z_0}\mathbf{Z_1}\mathbf{Z_2}\mathbf{Z_4}\mathbf{Z_5} + 0.0015\mathbf{Z_0}\mathbf{Z_1}\mathbf{X_2}\mathbf{X_3}\mathbf{Z_4}\mathbf{Z_5} + 0.0015\mathbf{Z_0}\mathbf{Z_1}\mathbf{Y_2}\mathbf{Y_3}\mathbf{Z_5} -0.0024\mathbf{Y_0}\mathbf{Z_1}\mathbf{Z_2}\mathbf{Y_3}\mathbf{Z_5} -0.0024\mathbf{X_0}\mathbf{Z_2}\mathbf{X_3}\mathbf{Z_4}\mathbf{Z_5} + 0.0847\mathbf{Z_0}\mathbf{Z_1}\mathbf{Z_3}\mathbf{Z_5} + 0.0090\mathbf{Y_0}\mathbf{Z_1}\mathbf{Y_2}\mathbf{Z_4}\mathbf{Z_5} + 0.0090\mathbf{X_0}\mathbf{X_2}\mathbf{Z_4}\mathbf{Z_5} + 0.0605\mathbf{Z_2}\mathbf{Z_3}\mathbf{Z_5} + 0.0042\mathbf{Z_0}\mathbf{Z_2}\mathbf{Y_3}\mathbf{Y_4}\mathbf{Z_5} + 0.0042\mathbf{Z_1}\mathbf{X_3}\mathbf{X_4} -0.0048\mathbf{Z_0}\mathbf{Z_1}\mathbf{X_3}\mathbf{Z_4}\mathbf{Z_5} -0.0048\mathbf{Z_1}\mathbf{Z_2}\mathbf{X_3}\mathbf{Z_4}\mathbf{Z_5} -0.0048\mathbf{Z_0}\mathbf{X_3} -0.0048\mathbf{Z_2}\mathbf{X_3} -0.0103\mathbf{Z_0}\mathbf{Z_1}\mathbf{X_2}\mathbf{Z_3}\mathbf{Z_5} -0.0103\mathbf{Z_1}\mathbf{X_2}\mathbf{Z_4}\mathbf{Z_5} -0.0103\mathbf{Z_0}\mathbf{X_2}\mathbf{Z_3}\mathbf{Z_4} -0.0103\mathbf{X_2} + 0.0035\mathbf{X_0}\mathbf{Z_1}\mathbf{Z_2}\mathbf{Z_3}\mathbf{Z_5} + 0.0035\mathbf{X_0}\mathbf{Z_4}\mathbf{Z_5} + 0.0035\mathbf{X_0}\mathbf{Z_2}\mathbf{Z_3}\mathbf{Z_4} + 0.0035\mathbf{X_0}\mathbf{Z_1} + 0.0042\mathbf{Z_1}\mathbf{X_3}\mathbf{X_4}\mathbf{Z_5} + 0.0042\mathbf{Z_0}\mathbf{Z_2}\mathbf{Y_3}\mathbf{Y_4} -0.0048\mathbf{Y_0}\mathbf{Y_1}\mathbf{X_3}\mathbf{Z_4}\mathbf{Z_5} + 0.0048\mathbf{Y_0}\mathbf{X_1}\mathbf{Y_3}\mathbf{Z_5} + 0.0048\mathbf{X_0}\mathbf{Y_1}\mathbf{Z_2}\mathbf{Y_3}\mathbf{Z_4} + 0.0048\mathbf{X_0}\mathbf{X_1}\mathbf{Z_2}\mathbf{X_3} -0.0103\mathbf{Y_0}\mathbf{Y_1}\mathbf{X_2}\mathbf{Z_3}\mathbf{Z_5} + 0.0103\mathbf{Y_0}\mathbf{X_1}\mathbf{Y_2}\mathbf{Z_3}\mathbf{Z_5} + 0.0103\mathbf{X_0}\mathbf{Y_1}\mathbf{Y_2} + 0.0103\mathbf{X_0}\mathbf{X_1}\mathbf{X_2} + 0.0035\mathbf{Z_0}\mathbf{X_1}\mathbf{Z_4}\mathbf{Z_5} -0.0035\mathbf{X_1}\mathbf{Z_4}\mathbf{Z_5} -0.0035\mathbf{X_1}\mathbf{Z_2}\mathbf{Z_3}\mathbf{Z_4} + 0.0035\mathbf{Z_0}\mathbf{X_1}\mathbf{Z_2}\mathbf{Z_3}\mathbf{Z_4} + 0.0127\mathbf{Y_3}\mathbf{Y_4}\mathbf{Z_5} + 0.0127\mathbf{Z_0}\mathbf{Z_1}\mathbf{Z_2}\mathbf{X_3}\mathbf{X_4} -0.0021\mathbf{Y_2}\mathbf{Z_3}\mathbf{Y_4}\mathbf{Z_5} -0.0021\mathbf{X_2}\mathbf{X_4}\mathbf{Z_5} -0.0021\mathbf{Z_0}\mathbf{Z_1}\mathbf{X_2}\mathbf{X_4} -0.0021\mathbf{Z_0}\mathbf{Z_1}\mathbf{Y_2}\mathbf{Z_3}\mathbf{Y_4} + 0.0078\mathbf{X_0}\mathbf{X_4}\mathbf{Z_5} + 0.0078\mathbf{Y_0}\mathbf{Z_1}\mathbf{Z_3}\mathbf{Y_4}\mathbf{Z_5} + 0.0078\mathbf{Y_0}\mathbf{Z_1}\mathbf{Z_2}\mathbf{Z_3}\mathbf{Y_4} + 0.0078\mathbf{X_0}\mathbf{Z_2}\mathbf{X_4} + 0.0090\mathbf{Z_2}\mathbf{X_3}\mathbf{X_4}\mathbf{Z_5} + 0.0090\mathbf{Z_0}\mathbf{Z_1}\mathbf{Y_3}\mathbf{Y_4} + 0.0066\mathbf{X_0}\mathbf{X_2}\mathbf{Y_3}\mathbf{Y_4}\mathbf{Z_5} + 0.0066\mathbf{Y_0}\mathbf{Z_1}\mathbf{Y_2}\mathbf{Y_3}\mathbf{Y_4}\mathbf{Z_5} -0.0066\mathbf{Y_0}\mathbf{Z_1}\mathbf{Y_2}\mathbf{X_3}\mathbf{X_4} -0.0066\mathbf{X_0}\mathbf{X_2}\mathbf{X_3}\mathbf{X_4} -0.0109\mathbf{Z_0}\mathbf{Z_1}\mathbf{X_3}\mathbf{X_4}\mathbf{Z_5} -0.0109\mathbf{Z_2}\mathbf{Y_3}\mathbf{Y_4} -0.2708\mathbf{Z_0}\mathbf{Z_1}\mathbf{Z_2}\mathbf{Z_3}\mathbf{Z_4} -0.1298\mathbf{Z_0}\mathbf{Z_2}\mathbf{Z_3} -0.1316\mathbf{Z_0}\mathbf{Z_2} + 0.0024\mathbf{Z_0}\mathbf{X_2}\mathbf{X_3} + 0.0024\mathbf{Z_0}\mathbf{Y_2}\mathbf{Y_3}\mathbf{Z_4} + 0.0006\mathbf{Y_0}\mathbf{Z_2}\mathbf{Y_3}\mathbf{Z_4} + 0.0006\mathbf{X_0}\mathbf{Z_1}\mathbf{Z_2}\mathbf{X_3} -0.1308\mathbf{Z_0}\mathbf{Z_3}\mathbf{Z_4} -0.0042\mathbf{Y_0}\mathbf{Y_2} -0.0042\mathbf{X_0}\mathbf{Z_1}\mathbf{X_2} -0.1298\mathbf{Z_1}\mathbf{Z_2}\mathbf{Z_3}\mathbf{Z_4} -0.0084\mathbf{X_0}\mathbf{X_1} + 0.0084\mathbf{Y_0}\mathbf{Y_1}\mathbf{Z_2}\mathbf{Z_3}\mathbf{Z_4} -0.0049\mathbf{Z_1}\mathbf{Z_2}\mathbf{Z_3}\mathbf{X_4} -0.0049\mathbf{Z_0}\mathbf{Z_1}\mathbf{Z_3}\mathbf{X_4} -0.0049\mathbf{Z_0}\mathbf{Z_1}\mathbf{X_4} -0.0049\mathbf{Z_1}\mathbf{Z_2}\mathbf{X_4} + 0.0035\mathbf{Z_1}\mathbf{Y_2}\mathbf{X_3}\mathbf{Y_4} + 0.0035\mathbf{Z_0}\mathbf{Z_1}\mathbf{Y_2}\mathbf{Y_3}\mathbf{X_4} + 0.0035\mathbf{Z_0}\mathbf{Z_1}\mathbf{X_2}\mathbf{Y_3}\mathbf{Y_4} -0.0035\mathbf{Z_1}\mathbf{X_2}\mathbf{X_3}\mathbf{X_4} -0.0049\mathbf{Y_0}\mathbf{Z_1}\mathbf{X_3}\mathbf{Y_4} -0.0049\mathbf{Y_0}\mathbf{Z_2}\mathbf{Y_3}\mathbf{X_4} -0.0049\mathbf{X_0}\mathbf{Z_1}\mathbf{Z_2}\mathbf{Y_3}\mathbf{Y_4} + 0.0049\mathbf{X_0}\mathbf{X_3}\mathbf{X_4} + 0.1298\mathbf{Z_1}\mathbf{Z_4} + 0.1316\mathbf{Z_1}\mathbf{Z_3}\mathbf{Z_4} + 0.0024\mathbf{Z_1}\mathbf{Y_2}\mathbf{Y_3}\mathbf{Z_4} + 0.0024\mathbf{Z_1}\mathbf{X_2}\mathbf{X_3} + 0.0006\mathbf{X_0}\mathbf{Z_1}\mathbf{X_3} + 0.0006\mathbf{Y_0}\mathbf{Y_3}\mathbf{Z_4} + 0.1308\mathbf{Z_1}\mathbf{Z_2} -0.0042\mathbf{X_0}\mathbf{Z_1}\mathbf{X_2}\mathbf{Z_3}\mathbf{Z_4} -0.0042\mathbf{Y_0}\mathbf{Y_2}\mathbf{Z_3}\mathbf{Z_4} + 0.1298\mathbf{Z_0} + 0.0049\mathbf{Y_0}\mathbf{X_1}\mathbf{Y_4} -0.0049\mathbf{Y_0}\mathbf{Y_1}\mathbf{Z_3}\mathbf{X_4} -0.0049\mathbf{Y_0}\mathbf{Y_1}\mathbf{X_4} + 0.0049\mathbf{Y_0}\mathbf{X_1}\mathbf{Z_3}\mathbf{Y_4} + 0.0035\mathbf{Y_0}\mathbf{X_1}\mathbf{X_2}\mathbf{Y_3}\mathbf{X_4} + 0.0035\mathbf{Y_0}\mathbf{Y_1}\mathbf{Y_2}\mathbf{Y_3}\mathbf{X_4} + 0.0035\mathbf{Y_0}\mathbf{Y_1}\mathbf{X_2}\mathbf{Y_3}\mathbf{Y_4} -0.0035\mathbf{Y_0}\mathbf{X_1}\mathbf{Y_2}\mathbf{Y_3}\mathbf{Y_4} + 0.0049\mathbf{Z_0}\mathbf{Y_1}\mathbf{X_3}\mathbf{Y_4} -0.0049\mathbf{Y_1}\mathbf{X_3}\mathbf{Y_4} + 0.0049\mathbf{Z_0}\mathbf{X_1}\mathbf{X_3}\mathbf{X_4} -0.0049\mathbf{X_1}\mathbf{X_3}\mathbf{X_4} + 0.1143\mathbf{Z_3} + 0.0105\mathbf{Y_2}\mathbf{Y_3} + 0.0105\mathbf{X_2}\mathbf{X_3}\mathbf{Z_4} -0.0351\mathbf{X_0}\mathbf{X_3}\mathbf{Z_4} -0.0351\mathbf{Y_0}\mathbf{Z_1}\mathbf{Y_3} + 0.0605\mathbf{Z_2}\mathbf{Z_4} + 0.0109\mathbf{X_0}\mathbf{X_2}\mathbf{Z_3} + 0.0109\mathbf{Y_0}\mathbf{Z_1}\mathbf{Y_2}\mathbf{Z_3} + 0.1141\mathbf{Z_0}\mathbf{Z_1}\mathbf{Z_4} + 0.0536\mathbf{Z_2}\mathbf{Z_3}\mathbf{Z_4} + 0.0044\mathbf{X_0}\mathbf{X_2} + 0.0044\mathbf{Y_0}\mathbf{Z_1}\mathbf{Y_2} + 0.0836\mathbf{Z_0}\mathbf{Z_1}\mathbf{Z_3}\mathbf{Z_4} + 0.0045\mathbf{X_0}\mathbf{Z_2}\mathbf{X_3} + 0.0045\mathbf{Y_0}\mathbf{Z_1}\mathbf{Z_2}\mathbf{Y_3}\mathbf{Z_4} + 0.0028\mathbf{Z_0}\mathbf{Z_1}\mathbf{Y_2}\mathbf{Y_3}\mathbf{Z_4} + 0.0028\mathbf{Z_0}\mathbf{Z_1}\mathbf{X_2}\mathbf{X_3} + 0.0539\mathbf{Z_0}\mathbf{Z_1}\mathbf{Z_2}$

\subsection{\label{app:beh2} $\rm BeH_2$ molecule}

$H_{\rm BeH_2} =  -13.4710 - 0.0610\mathbf{Z_0}\mathbf{Z_1}\mathbf{Z_2}\mathbf{Z_4}\mathbf{Z_5} - 0.0159\mathbf{Z_4}\mathbf{X_5} + 0.0159\mathbf{Z_1}\mathbf{Z_3}\mathbf{X_5} - 0.0787\mathbf{Z_0}\mathbf{Z_1}\mathbf{Z_2}\mathbf{Z_4} + 0.0375\mathbf{Z_0}\mathbf{Z_1}\mathbf{Z_2}\mathbf{X_3}\mathbf{Y_4}\mathbf{Y_5} - 0.0375\mathbf{X_3}\mathbf{Y_4}\mathbf{Y_5} - 0.4753\mathbf{Z_0}\mathbf{Z_2}\mathbf{Z_3}\mathbf{Z_4} - 0.2865\mathbf{Z_0}\mathbf{Z_2}\mathbf{Z_3}\mathbf{Z_5} - 0.4765\mathbf{Z_4} - 0.0610\mathbf{Z_3}\mathbf{Z_4} + 0.0159\mathbf{Y_0}\mathbf{Y_1}\mathbf{X_3} - 0.0159\mathbf{Y_0}\mathbf{X_1}\mathbf{Y_3}\mathbf{Z_4} - 0.0787\mathbf{Z_2} - 0.0375\mathbf{X_0}\mathbf{X_2}\mathbf{Z_3}\mathbf{Z_4} - 0.0375\mathbf{Y_0}\mathbf{Z_1}\mathbf{Y_2}\mathbf{Z_3}\mathbf{Z_4} - 0.2865\mathbf{Z_1} - 0.4765\mathbf{Z_0}\mathbf{Z_1} + 0.0581\mathbf{Z_5} + 0.0099\mathbf{X_3}\mathbf{X_4}\mathbf{X_5} - 0.0099\mathbf{Z_0}\mathbf{Z_1}\mathbf{Z_2}\mathbf{X_3}\mathbf{X_4}\mathbf{X_5} + 0.1668\mathbf{Z_1}\mathbf{Z_3}\mathbf{Z_5} + 0.1500\mathbf{Z_1}\mathbf{Z_3}\mathbf{Z_4} + 0.0876\mathbf{Z_0}\mathbf{Z_1}\mathbf{Z_2}\mathbf{Z_5} + 0.1947\mathbf{Z_0}\mathbf{Z_1}\mathbf{Z_2}\mathbf{Z_3}\mathbf{Z_5} + 0.0030\mathbf{X_0}\mathbf{X_1}\mathbf{Z_2}\mathbf{X_3}\mathbf{Z_4}\mathbf{Z_5} + 0.0030\mathbf{X_0}\mathbf{Y_1}\mathbf{Z_2}\mathbf{Y_3}\mathbf{Z_5} + 0.0982\mathbf{Z_0}\mathbf{Z_1}\mathbf{Z_4}\mathbf{Z_5} + 0.0002\mathbf{Y_0}\mathbf{Z_1}\mathbf{Y_2}\mathbf{Z_3}\mathbf{Z_5} + 0.0002\mathbf{X_0}\mathbf{X_2}\mathbf{Z_3}\mathbf{Z_5} + 0.1908\mathbf{Z_0}\mathbf{Z_2}\mathbf{Z_4}\mathbf{Z_5} + 0.1019\mathbf{Z_2}\mathbf{Z_4}\mathbf{Z_5} + 0.0125\mathbf{Z_0}\mathbf{Z_1}\mathbf{Z_2}\mathbf{X_5} - 0.0125\mathbf{Z_0}\mathbf{Z_2}\mathbf{Z_3}\mathbf{Z_4}\mathbf{X_5} - 0.0196\mathbf{Z_0}\mathbf{Z_2}\mathbf{Y_3}\mathbf{Y_4} - 0.0159\mathbf{X_3}\mathbf{X_4}\mathbf{Z_5} - 0.0038\mathbf{Z_0}\mathbf{Z_1}\mathbf{Z_2}\mathbf{X_3}\mathbf{X_4}\mathbf{Z_5} + 0.0038\mathbf{Z_1}\mathbf{Y_3}\mathbf{Y_4}\mathbf{Z_5} + 0.0159\mathbf{Z_0}\mathbf{Z_2}\mathbf{Y_3}\mathbf{Y_4}\mathbf{Z_5} + 0.0196\mathbf{X_3}\mathbf{X_4} - 0.0401\mathbf{Y_2}\mathbf{Y_3}\mathbf{Z_4}\mathbf{Z_5} - 0.0401\mathbf{X_2}\mathbf{X_3}\mathbf{Z_5} + 0.0401\mathbf{Y_2}\mathbf{Y_3}\mathbf{Z_4} + 0.0401\mathbf{X_2}\mathbf{X_3} - 0.0101\mathbf{X_0}\mathbf{X_3}\mathbf{Z_5} - 0.0101\mathbf{Y_0}\mathbf{Z_1}\mathbf{Y_3}\mathbf{Z_4}\mathbf{Z_5} + 0.0101\mathbf{X_0}\mathbf{X_3} + 0.0101\mathbf{Y_0}\mathbf{Z_1}\mathbf{Y_3}\mathbf{Z_4} + 0.0231\mathbf{Y_0}\mathbf{Y_1}\mathbf{X_2}\mathbf{Z_3}\mathbf{Z_4}\mathbf{Z_5} - 0.0231\mathbf{Y_0}\mathbf{X_1}\mathbf{Y_2}\mathbf{Z_3}\mathbf{Z_4}\mathbf{Z_5} - 0.0231\mathbf{Y_0}\mathbf{Y_1}\mathbf{X_2}\mathbf{Z_3}\mathbf{Z_4} + 0.0231\mathbf{Y_0}\mathbf{X_1}\mathbf{Y_2}\mathbf{Z_3}\mathbf{Z_4} + 0.0352\mathbf{Z_0}\mathbf{X_1}\mathbf{Z_5} - 0.0352\mathbf{X_1}\mathbf{Z_5} - 0.0352\mathbf{Z_0}\mathbf{X_1} + 0.0352\mathbf{X_1} + 0.0201\mathbf{Z_0}\mathbf{Z_2}\mathbf{Z_3}\mathbf{X_5} - 0.0201\mathbf{Z_0}\mathbf{Z_1}\mathbf{Z_2}\mathbf{Z_4}\mathbf{X_5} - 0.0120\mathbf{Z_2}\mathbf{Y_3}\mathbf{Z_4}\mathbf{Y_5} + 0.0120\mathbf{Z_0}\mathbf{Y_3}\mathbf{Z_4}\mathbf{Y_5} - 0.0120\mathbf{Z_1}\mathbf{Z_2}\mathbf{X_3}\mathbf{Z_4}\mathbf{X_5} + 0.0120\mathbf{Z_0}\mathbf{Z_1}\mathbf{X_3}\mathbf{Z_4}\mathbf{X_5} - 0.0117\mathbf{X_0}\mathbf{Y_1}\mathbf{Y_5} + 0.0117\mathbf{Y_0}\mathbf{X_1}\mathbf{Z_2}\mathbf{Z_3}\mathbf{Z_4}\mathbf{Y_5} - 0.0117\mathbf{X_0}\mathbf{X_1}\mathbf{Z_3}\mathbf{X_5} - 0.0117\mathbf{Y_0}\mathbf{Y_1}\mathbf{Z_2}\mathbf{Z_4}\mathbf{X_5} + 0.0065\mathbf{X_5} - 0.0065\mathbf{Z_1}\mathbf{Z_3}\mathbf{Z_4}\mathbf{X_5} + 0.0030\mathbf{Z_3}\mathbf{X_5} - 0.0030\mathbf{Z_1}\mathbf{Z_4}\mathbf{X_5} - 0.0408\mathbf{Y_0}\mathbf{Y_1}\mathbf{X_3}\mathbf{Z_4}\mathbf{X_5} + 0.0408\mathbf{Y_0}\mathbf{X_1}\mathbf{Y_3}\mathbf{X_5} - 0.0106\mathbf{Z_2}\mathbf{Z_4}\mathbf{X_5} + 0.0106\mathbf{Z_1}\mathbf{Z_2}\mathbf{Z_3}\mathbf{X_5} + 0.0193\mathbf{X_0}\mathbf{X_2}\mathbf{Z_3}\mathbf{X_5} + 0.0193\mathbf{Y_0}\mathbf{Z_1}\mathbf{Y_2}\mathbf{Z_3}\mathbf{X_5} - 0.0193\mathbf{X_0}\mathbf{Z_1}\mathbf{X_2}\mathbf{Z_4}\mathbf{X_5} - 0.0193\mathbf{Y_0}\mathbf{Y_2}\mathbf{Z_4}\mathbf{X_5} - 0.0058\mathbf{Z_0}\mathbf{Z_1}\mathbf{Z_4}\mathbf{X_5} + 0.0058\mathbf{Z_0}\mathbf{Z_3}\mathbf{X_5} + 0.0101\mathbf{Y_2}\mathbf{Z_3}\mathbf{Y_4}\mathbf{X_5} + 0.0101\mathbf{X_2}\mathbf{X_4}\mathbf{X_5} - 0.0101\mathbf{Z_0}\mathbf{Z_1}\mathbf{X_2}\mathbf{Z_3}\mathbf{Y_4}\mathbf{Y_5} + 0.0101\mathbf{Z_0}\mathbf{Z_1}\mathbf{Y_2}\mathbf{X_4}\mathbf{Y_5} + 0.0143\mathbf{X_0}\mathbf{X_4}\mathbf{X_5} + 0.0143\mathbf{Y_0}\mathbf{Z_1}\mathbf{Z_3}\mathbf{Y_4}\mathbf{X_5} + 0.0143\mathbf{Y_0}\mathbf{Z_1}\mathbf{Z_2}\mathbf{X_4}\mathbf{Y_5} - 0.0143\mathbf{X_0}\mathbf{Z_2}\mathbf{Z_3}\mathbf{Y_4}\mathbf{Y_5} + 0.0155\mathbf{Y_0}\mathbf{Y_1}\mathbf{X_2}\mathbf{Y_3}\mathbf{Y_4}\mathbf{X_5} - 0.0155\mathbf{Y_0}\mathbf{X_1}\mathbf{Y_2}\mathbf{Y_3}\mathbf{Y_4}\mathbf{X_5} - 0.0155\mathbf{X_0}\mathbf{X_1}\mathbf{Y_2}\mathbf{Y_3}\mathbf{Y_4}\mathbf{Y_5} + 0.0155\mathbf{X_0}\mathbf{Y_1}\mathbf{X_2}\mathbf{Y_3}\mathbf{Y_4}\mathbf{Y_5} - 0.0124\mathbf{Z_0}\mathbf{X_1}\mathbf{X_3}\mathbf{X_4}\mathbf{X_5} + 0.0124\mathbf{X_1}\mathbf{X_3}\mathbf{X_4}\mathbf{X_5} - 0.0124\mathbf{Y_1}\mathbf{Z_2}\mathbf{X_3}\mathbf{X_4}\mathbf{Y_5} + 0.0124\mathbf{Z_0}\mathbf{Y_1}\mathbf{Z_2}\mathbf{X_3}\mathbf{X_4}\mathbf{Y_5} + 0.1672\mathbf{Z_1}\mathbf{Z_3} + 0.0767\mathbf{Z_1}\mathbf{Z_3}\mathbf{Z_4}\mathbf{Z_5} + 0.1693\mathbf{Z_0}\mathbf{Z_1}\mathbf{Z_2} + 0.0982\mathbf{Z_0}\mathbf{Z_1}\mathbf{Z_2}\mathbf{Z_3} - 0.0106\mathbf{X_0}\mathbf{X_1}\mathbf{Z_2}\mathbf{X_3}\mathbf{Z_4} - 0.0106\mathbf{X_0}\mathbf{Y_1}\mathbf{Z_2}\mathbf{Y_3} + 0.2199\mathbf{Z_0}\mathbf{Z_1}\mathbf{Z_4} + 0.0045\mathbf{Y_0}\mathbf{Z_1}\mathbf{Y_2}\mathbf{Z_3} + 0.0045\mathbf{X_0}\mathbf{X_2}\mathbf{Z_3} + 0.0982\mathbf{Z_0}\mathbf{Z_2}\mathbf{Z_4} + 0.2123\mathbf{Z_2}\mathbf{Z_4} - 0.0304\mathbf{Z_1}\mathbf{Y_3}\mathbf{X_4}\mathbf{Y_5} + 0.0304\mathbf{Z_0}\mathbf{Z_2}\mathbf{Y_3}\mathbf{X_4}\mathbf{Y_5} + 0.0033\mathbf{Z_0}\mathbf{Z_1}\mathbf{X_2}\mathbf{Z_4}\mathbf{X_5} - 0.0033\mathbf{Z_1}\mathbf{X_2}\mathbf{Z_3}\mathbf{X_5} - 0.0033\mathbf{Z_0}\mathbf{X_2}\mathbf{Z_3}\mathbf{Z_4}\mathbf{X_5} + 0.0033\mathbf{X_2}\mathbf{X_5} + 0.0034\mathbf{X_0}\mathbf{Z_1}\mathbf{Z_2}\mathbf{Z_4}\mathbf{X_5} - 0.0034\mathbf{X_0}\mathbf{Z_3}\mathbf{X_5} - 0.0034\mathbf{X_0}\mathbf{Z_2}\mathbf{Z_3}\mathbf{Z_4}\mathbf{X_5} + 0.0034\mathbf{X_0}\mathbf{Z_1}\mathbf{X_5} - 0.0177\mathbf{Z_1}\mathbf{Y_3}\mathbf{Y_4}\mathbf{X_5} + 0.0177\mathbf{Z_0}\mathbf{Z_2}\mathbf{Y_3}\mathbf{Y_4}\mathbf{X_5} + 0.0231\mathbf{Z_0}\mathbf{Z_1}\mathbf{X_2}\mathbf{Y_3}\mathbf{Z_4}\mathbf{Y_5} - 0.0231\mathbf{Z_0}\mathbf{Z_1}\mathbf{Y_2}\mathbf{X_3}\mathbf{Y_5} + 0.0231\mathbf{Z_0}\mathbf{X_2}\mathbf{X_3}\mathbf{X_5} + 0.0231\mathbf{Z_0}\mathbf{Y_2}\mathbf{Y_3}\mathbf{Z_4}\mathbf{X_5} - 0.0155\mathbf{Y_0}\mathbf{Z_1}\mathbf{Z_2}\mathbf{X_3}\mathbf{Y_5} + 0.0155\mathbf{X_0}\mathbf{Z_2}\mathbf{Y_3}\mathbf{Z_4}\mathbf{Y_5} + 0.0155\mathbf{Y_0}\mathbf{Z_2}\mathbf{Y_3}\mathbf{Z_4}\mathbf{X_5} + 0.0155\mathbf{X_0}\mathbf{Z_1}\mathbf{Z_2}\mathbf{X_3}\mathbf{X_5} - 0.0215\mathbf{X_0}\mathbf{X_1}\mathbf{Y_2}\mathbf{Z_3}\mathbf{Z_4}\mathbf{Y_5} + 0.0215\mathbf{X_0}\mathbf{Y_1}\mathbf{X_2}\mathbf{Z_3}\mathbf{Z_4}\mathbf{Y_5} + 0.0215\mathbf{X_0}\mathbf{Y_1}\mathbf{Y_2}\mathbf{X_5} + 0.0215\mathbf{X_0}\mathbf{X_1}\mathbf{X_2}\mathbf{X_5} + 0.0231\mathbf{Y_1}\mathbf{Z_2}\mathbf{Y_5} - 0.0231\mathbf{Z_0}\mathbf{Y_1}\mathbf{Z_2}\mathbf{Y_5} + 0.0231\mathbf{X_1}\mathbf{Z_2}\mathbf{Z_3}\mathbf{Z_4}\mathbf{X_5} - 0.0231\mathbf{Z_0}\mathbf{X_1}\mathbf{Z_2}\mathbf{Z_3}\mathbf{Z_4}\mathbf{X_5} + 0.0002\mathbf{Z_0}\mathbf{Z_1}\mathbf{Z_2}\mathbf{Y_3}\mathbf{X_4}\mathbf{Y_5} - 0.0002\mathbf{Y_3}\mathbf{X_4}\mathbf{Y_5} - 0.0193\mathbf{X_0}\mathbf{X_1}\mathbf{Z_2}\mathbf{Y_4}\mathbf{Y_5} + 0.0193\mathbf{X_0}\mathbf{Y_1}\mathbf{Z_2}\mathbf{Z_3}\mathbf{X_4}\mathbf{Y_5} - 0.0193\mathbf{Y_0}\mathbf{Y_1}\mathbf{Y_4}\mathbf{Y_5} - 0.0193\mathbf{Y_0}\mathbf{X_1}\mathbf{Z_3}\mathbf{X_4}\mathbf{Y_5} + 0.0045\mathbf{Z_0}\mathbf{Z_1}\mathbf{X_3}\mathbf{Y_4}\mathbf{Y_5} - 0.0045\mathbf{Z_2}\mathbf{X_3}\mathbf{Y_4}\mathbf{Y_5} + 0.0429\mathbf{Y_0}\mathbf{Z_1}\mathbf{Y_2}\mathbf{Y_3}\mathbf{X_4}\mathbf{Y_5} + 0.0429\mathbf{X_0}\mathbf{X_2}\mathbf{Y_3}\mathbf{X_4}\mathbf{Y_5} + 0.0054\mathbf{Z_0}\mathbf{Z_2}\mathbf{X_3}\mathbf{Y_4}\mathbf{Y_5} - 0.0054\mathbf{Z_1}\mathbf{X_3}\mathbf{Y_4}\mathbf{Y_5} + 0.1659\mathbf{Z_4}\mathbf{Z_5} + 0.1864\mathbf{Z_0}\mathbf{Z_2}\mathbf{Z_3} + 0.1668\mathbf{Z_0}\mathbf{Z_2} - 0.0201\mathbf{X_0}\mathbf{Y_1}\mathbf{Z_2}\mathbf{Y_3}\mathbf{Z_4} - 0.0201\mathbf{X_0}\mathbf{X_1}\mathbf{Z_2}\mathbf{X_3} + 0.1672\mathbf{Z_0}\mathbf{Z_3}\mathbf{Z_4} - 0.0304\mathbf{Y_0}\mathbf{Y_2} - 0.0304\mathbf{X_0}\mathbf{Z_1}\mathbf{X_2} + 0.1659\mathbf{Z_0}\mathbf{Z_1}\mathbf{Z_2}\mathbf{Z_3}\mathbf{Z_4} + 0.1864\mathbf{Z_1}\mathbf{Z_2}\mathbf{Z_3}\mathbf{Z_4} + 0.0117\mathbf{Z_0}\mathbf{X_3}\mathbf{Z_4}\mathbf{Z_5} - 0.0117\mathbf{Z_2}\mathbf{X_3}\mathbf{Z_4}\mathbf{Z_5} - 0.0117\mathbf{Z_0}\mathbf{X_3} + 0.0117\mathbf{Z_2}\mathbf{X_3} - 0.0125\mathbf{Y_0}\mathbf{Y_1}\mathbf{Z_2}\mathbf{Z_3}\mathbf{Z_5} - 0.0125\mathbf{X_0}\mathbf{X_1}\mathbf{Z_4}\mathbf{Z_5} + 0.0125\mathbf{Y_0}\mathbf{Y_1}\mathbf{Z_2}\mathbf{Z_3}\mathbf{Z_4} + 0.0125\mathbf{X_0}\mathbf{X_1} - 0.0034\mathbf{Z_0}\mathbf{X_2}\mathbf{Y_3}\mathbf{Y_4}\mathbf{Z_5} - 0.0034\mathbf{X_2}\mathbf{X_3}\mathbf{X_4}\mathbf{Z_5} + 0.0034\mathbf{Y_2}\mathbf{X_3}\mathbf{Y_4}\mathbf{Z_5} - 0.0034\mathbf{Z_0}\mathbf{Y_2}\mathbf{Y_3}\mathbf{X_4}\mathbf{Z_5} - 0.0042\mathbf{X_0}\mathbf{Z_2}\mathbf{Y_3}\mathbf{Y_4}\mathbf{Z_5} - 0.0042\mathbf{X_0}\mathbf{Z_1}\mathbf{X_3}\mathbf{X_4}\mathbf{Z_5} + 0.0042\mathbf{Y_0}\mathbf{X_3}\mathbf{Y_4}\mathbf{Z_5} - 0.0042\mathbf{Y_0}\mathbf{Z_1}\mathbf{Z_2}\mathbf{Y_3}\mathbf{X_4}\mathbf{Z_5} + 0.0699\mathbf{Z_0}\mathbf{Z_2}\mathbf{Z_3}\mathbf{Z_4}\mathbf{Z_5} + 0.0982\mathbf{Z_0}\mathbf{Z_3}\mathbf{Z_5} + 0.0054\mathbf{Y_0}\mathbf{Y_2}\mathbf{Z_4}\mathbf{Z_5} + 0.0054\mathbf{X_0}\mathbf{Z_1}\mathbf{X_2}\mathbf{Z_4}\mathbf{Z_5} + 0.1043\mathbf{Z_1}\mathbf{Z_2}\mathbf{Z_3}\mathbf{Z_5} + 0.0352\mathbf{Z_0}\mathbf{X_2}\mathbf{X_4}\mathbf{Z_5} + 0.0352\mathbf{Z_0}\mathbf{Y_2}\mathbf{Z_3}\mathbf{Y_4}\mathbf{Z_5} + 0.0352\mathbf{Y_2}\mathbf{Z_3}\mathbf{Y_4} + 0.0352\mathbf{X_2}\mathbf{X_4} + 0.0124\mathbf{Y_0}\mathbf{Z_2}\mathbf{Z_3}\mathbf{Y_4}\mathbf{Z_5} + 0.0124\mathbf{X_0}\mathbf{Z_1}\mathbf{Z_2}\mathbf{X_4}\mathbf{Z_5} + 0.0124\mathbf{X_0}\mathbf{X_4} + 0.0124\mathbf{Y_0}\mathbf{Z_1}\mathbf{Z_3}\mathbf{Y_4} + 0.0231\mathbf{X_0}\mathbf{Y_1}\mathbf{Y_2}\mathbf{X_3}\mathbf{X_4}\mathbf{Z_5} + 0.0231\mathbf{X_0}\mathbf{X_1}\mathbf{X_2}\mathbf{X_3}\mathbf{X_4}\mathbf{Z_5} + 0.0231\mathbf{Y_0}\mathbf{Y_1}\mathbf{X_2}\mathbf{Y_3}\mathbf{Y_4} - 0.0231\mathbf{Y_0}\mathbf{X_1}\mathbf{Y_2}\mathbf{Y_3}\mathbf{Y_4} - 0.0344\mathbf{X_1}\mathbf{Z_2}\mathbf{Y_3}\mathbf{Y_4}\mathbf{Z_5} + 0.0344\mathbf{Z_0}\mathbf{X_1}\mathbf{Z_2}\mathbf{Y_3}\mathbf{Y_4}\mathbf{Z_5} - 0.0344\mathbf{Z_0}\mathbf{X_1}\mathbf{X_3}\mathbf{X_4} + 0.0344\mathbf{X_1}\mathbf{X_3}\mathbf{X_4} + 0.1019\mathbf{Z_3} + 0.0058\mathbf{Y_0}\mathbf{Y_1}\mathbf{X_3}\mathbf{Z_4} - 0.0058\mathbf{Y_0}\mathbf{X_1}\mathbf{Y_3} + 0.1043\mathbf{Z_1}\mathbf{Z_4} + 0.0581\mathbf{Z_2}\mathbf{Z_3}\mathbf{Z_4} + 0.0099\mathbf{X_0}\mathbf{X_2} + 0.0099\mathbf{Y_0}\mathbf{Z_1}\mathbf{Y_2} + 0.0876\mathbf{Z_0}\mathbf{Z_1}\mathbf{Z_3}\mathbf{Z_4} - 0.0125\mathbf{Y_0}\mathbf{Y_1}\mathbf{Z_2}\mathbf{X_3} + 0.0125\mathbf{Y_0}\mathbf{X_1}\mathbf{Z_2}\mathbf{Y_3}\mathbf{Z_4} - 0.0196\mathbf{Z_0}\mathbf{X_1}\mathbf{Y_2}\mathbf{Y_3}\mathbf{Z_4} + 0.0159\mathbf{X_1}\mathbf{Y_2}\mathbf{Y_3}\mathbf{Z_4} + 0.0038\mathbf{Z_0}\mathbf{Y_1}\mathbf{X_2}\mathbf{Y_3}\mathbf{Z_4} + 0.0038\mathbf{Y_1}\mathbf{Y_2}\mathbf{X_3} - 0.0159\mathbf{Z_0}\mathbf{X_1}\mathbf{X_2}\mathbf{X_3} + 0.0196\mathbf{X_1}\mathbf{X_2}\mathbf{X_3} + 0.0065\mathbf{X_0}\mathbf{X_1}\mathbf{X_3} + 0.0065\mathbf{X_0}\mathbf{Y_1}\mathbf{Y_3}\mathbf{Z_4} + 0.0767\mathbf{Z_1}\mathbf{Z_2} + 0.0177\mathbf{X_0}\mathbf{Z_1}\mathbf{X_2}\mathbf{Z_3}\mathbf{Z_4} + 0.0177\mathbf{Y_0}\mathbf{Y_2}\mathbf{Z_3}\mathbf{Z_4} + 0.0699\mathbf{Z_0}$
    
\end{widetext}


\bibliography{references}

\end{document}